\documentclass[11pt]{article}
\usepackage{amsmath,bm}
\usepackage{amssymb}
\usepackage{geometry}
\usepackage{graphicx}
\usepackage{hyperref}
\usepackage{subfig}
\usepackage{mciteplus}

\def\gtwid{\mathrel{\raise.3ex\hbox{$>$\kern-.75em\lower1ex\hbox{$\sim
$}}}}
\def\vio{\mathrel{\hbox{$E$\kern-.60em\hbox{$/
$}}}}
\newcommand{\newc}{\newcommand*}

\long\def\begincomment#1\endcomment{%
        \begingroup\sf\baselineskip12pt#1\endgroup}

\newc{\etal}{\textrm{et al.}} 
\newc{\eg}{\textrm{e.g.}} 
\newc{\ie}{\textrm{i.e.}}
\newc{\etc}{\textrm{etc}}
\newc\vs{\textrm{vs.}}
\newc{\cl}{\rm {C.L.}}

\newc{\ev}{\ensuremath{\,\mathrm{eV}}}
\newc{\kev}{\ensuremath{\,\mathrm{keV}}}
\newc{\mev}{\ensuremath{\,\mathrm{MeV}}}
\newc{\gev}{\ensuremath{\,\mathrm{GeV}}}
\newc{\tev}{\ensuremath{\,\mathrm{TeV}}}
\newc{\MeV}{\mev} 
\newc{\TeV}{\tev}
\newc{\invpb}{\ensuremath{/\text{pb}}}
\newc{\invfb}{\ensuremath{/\text{fb}}}
\newc\nb{\ensuremath{\,\mathrm{nb}}} \newc\pb{\ensuremath{\,\mathrm{pb}}} \newc\fb{\ensuremath{\,\mathrm{fb}}}
\newc\pc{\ensuremath{\,\mathrm{pc}}}
\newc\kpc{\ensuremath{\,\mathrm{kpc}}}
\newc\mpc{\ensuremath{\,\mathrm{Mpc}}}
\newc\ps{\ensuremath{\,\mathrm{ps}}} 
\newc\cmeter{\ensuremath{\,\mathrm{cm}}} 
\newc\meter{\ensuremath{\,\mathrm{m}}} 
\newc\kmeter{\ensuremath{\,\mathrm{km}}}
\newc\second{\ensuremath{\,\mathrm{s}}}
\newc\msecond{\ensuremath{\,\mathrm{ms}}}
\newc\nsecond{\ensuremath{\,\mathrm{ns}}}
\newc\psecond{\ensuremath{\,\mathrm{ps}}}

\newc{\chisqmin}{\ensuremath{\chi^2_{\mathrm{min}}}}
\newc{\Delchisq}{\ensuremath{\Delta\chi^2}}
\newc{\chisq}{\ensuremath{\chi^2}}
\newc{\like}{\ensuremath{\mathcal{L}}}

\newc\lsim{\ensuremath{\mathrel{\rlap{\lower4pt\hbox{\hskip1pt$\sim$}}\raise1pt\hbox{$<$}}}}
\newc\gsim{\ensuremath{\mathrel{\rlap{\lower4pt\hbox{\hskip1pt$\sim$}}\raise1pt\hbox{$>$}}}}
\newc{\VEV}[1]{\ensuremath{\langle #1 \rangle}}
\newc{\dl}{\ensuremath{\stackrel{\leftarrow}{D}}}
\newc{\dr}{\ensuremath{\stackrel{\rightarrow}{D}}}

\newc{\bcenter}{\begin{center}}    \newc{\ecenter}{\end{center}}
\newc{\bfl}{\begin{flushleft}}    \newc{\efl}{\end{flushleft}}
\newc{\bfr}{\begin{flushright}}    \newc{\efr}{\end{flushright}}

\newc{\bi}{\begin{itemize}}
\newc{\ei}{\end{itemize}}
\newc{\bed}{\begin{description}}
\newc{\eed}{\end{description}}
\newc{\ben}{\begin{enumerate}}
\newc{\een}{\end{enumerate}}

\newc{\be}{\begin{equation}}
\newc{\ee}{\end{equation}}
\newc{\bea}{\begin{eqnarray}}
\newc{\eea}{\end{eqnarray}}
\newc{\bfle}{\begin{flalign}}
\newc{\efle}{\end{flalign}}
\newc{\ra}{\rightarrow}

\newc{\alphas}{\ensuremath{\alpha_s}}
\newc{\alphatwo}{\ensuremath{\alpha_2}}
\newc{\alphaone}{\ensuremath{\alpha_1}}
\newc{\alphai}[1]{\ensuremath{\alpha_{#1}}}
\newc{\alphaem}{\ensuremath{\alpha_{\mathrm{em}}}}
\newc{\alphaeff}{\ensuremath{\alpha_{\mathrm{eff}}}}
\newc{\sineff}{\ensuremath{\sin \theta_{\mathrm{eff}}}}
\newc{\sinsqeff}{\ensuremath{\sin^2 \theta_{\mathrm{eff}}}}
\newc{\dalphahad}{\ensuremath{\Delta \alpha_{\mathrm{had}}}}
\newc{\yt}{\ensuremath{h_t}} \newc{\yb}{\ensuremath{h_b}} \newc{\ytau}{\ensuremath{h_{\tau}}}
\newc\mz{\ensuremath{M_Z}} 
\newc\mw{\ensuremath{m_W}}
\newc\mZ{\mz}        \newc\mW{\mw}
\newc\mhsm{\ensuremath{ m_{H_{\mathrm{SM}}}}}
\newc{\mtop}{\ensuremath{ m_t}}               \newc{\mtpole}{\ensuremath{ M_t}}
\newc{\mbottom}{\ensuremath{ m_b}} 
\newc{\mtau}{\ensuremath{ m_{\tau}}}
\newc{\mt}{\mtpole}
\newc{\mb}{\mbottom} 
\newc{\rtwogg}{\ensuremath{R_{h_2}(\gamma\gamma)}}
\newc{\rtwozz}{\ensuremath{R_{h_2}(ZZ)}}
\newc{\ronegg}{\ensuremath{R_{h_1}(\gamma\gamma)}}
\newc{\ronezz}{\ensuremath{R_{h_1}(ZZ)}}
\newc{\rsiggg}{\ensuremath{R_{h_\textrm{sig}}(\gamma\gamma)}}
\newc{\rsigzz}{\ensuremath{R_{h_\textrm{sig}}(ZZ)}}
\newc{\llbar}{\ensuremath{\ell\bar{\ell}}}
\newc{\tauptaum}{\ensuremath{ \tau^+\tau^-}}
\newc{\qqbar}{\ensuremath{ q\bar{q}}} \newc{\ppbar}{\ensuremath{ p\bar{p}}}
\newc{\bbbar}{\ensuremath{ b\bar{b}}} \newc{\ttbar}{\ensuremath{ t\bar{t}}}
\newc{\ffbar}{\ensuremath{ f\bar{f}}} \newc{\tautaubar}{\ensuremath{ \tau\bar{\tau}}}

\newc{\mchi}{\ensuremath{m_\neutone}}
\newc{\squark}{\ensuremath{\tilde{q}}}
\newc{\slepton}{\ensuremath{\tilde{l}}}
\newc{\gluino}{\ensuremath{\tilde{g}}} 
\newc{\mgluino}{\ensuremath{{m_{\gluino}}}}
\newc{\wino}{\ensuremath{\tilde{W}}} 
\newc{\mwino}{\ensuremath{{m_{\wino}}}}
\newc{\tone}{\ensuremath{{\tilde{t}_1}}}
\newc{\Hone}{\ensuremath{{\tilde{H}_{1}}}}
\newc{\Htwo}{\ensuremath{{\tilde{H}_{2}}}}
\newc{\Hhtwo}{\ensuremath{{H_{2}}}}
\newc{\qli}{\ensuremath{{\tilde{Q}_{i}}}}
\newc{\uri}{\ensuremath{{\tilde{u}_{i}}}}
\newc{\dri}{\ensuremath{{\tilde{d}_{i}}}}
\newc{\lli}{\ensuremath{{\tilde{L}_{i}}}}
\newc{\eri}{\ensuremath{{\tilde{e}_{i}}}}

\newc{\sthw}{\ensuremath{ \sin\theta_W}}              \newc{\cthw}{\ensuremath{\cos\theta_W}}
\newc{\tanthw}{\ensuremath{ \tan\theta_W}}              \newc{\cotthw}{\ensuremath{\cot\theta_W}}
\newc{\ssqthw}{\ensuremath{\sin^2 \theta_W}}
\newc{\msbar}{\ensuremath{\overline{MS}}} \newc{\drbar}{\ensuremath{\overline{DR}}}
\newc{\mtmtsmmsbar}{\ensuremath{ m_t(m_t)^{\msbar}_{{\mathrm{SM}}}}}
\newc{\mtmtsmdrbar}{\ensuremath{ m_t(m_t)^{\drbar}_{{\mathrm{SM}}}}}
\newc{\mtmtmssmdrbar}{\ensuremath{ m_t(m_t)^{\drbar}_{{\mathrm{SUSY}}}}}
\newc{\mbmbmsbar}{\ensuremath{ m_b(m_b)^{\msbar} }}
\newc{\mbmbsmmsbar}{\ensuremath{ m_b(m_b)^{\msbar}_{{\mathrm{SM}}}}}
\newc{\mbmzsmmsbar}{\ensuremath{ m_b(\mz)^{\msbar}_{{\mathrm{SM}}}}}
\newc{\mbmzsmdrbar}{\ensuremath{ m_b(\mz)^{\drbar}_{{\mathrm{SM}}}}}
\newc{\mbmzmssmdrbar}{\ensuremath{ m_b(\mz)^{\drbar}_{{\mathrm{SUSY}}}}}
\newc{\mtaumzsmmsbar}{\ensuremath{ m_{\tau}(\mz)^{\msbar}_{{\mathrm{SM}}}}}
\newc{\mtaumzsmdrbar}{\ensuremath{ m_{\tau}(\mz)^{\drbar}_{{\mathrm{SM}}}}}
\newc{\mtaumzmssmdrbar}{\ensuremath{ m_{\tau}(\mz)^{\drbar}_{{\mathrm{SUSY}}}}}
\newc{\alphasmzms}{\ensuremath{\alpha_s(M_Z)^{\overline{MS}}}}
\newc{\alphaimzms}[1]{\ensuremath{\alpha_{#1}(M_Z)^{\overline{MS}}}}
\newc{\alphaemmz}{\ensuremath{\alpha_{\mathrm{em}}(M_Z)^{\overline{MS}}}}

\newc{\mzero}{\ensuremath{{m_0}}}
\newc{\mhalf}{\ensuremath{ m_{1/2}}}
\newc{\tanb}{\ensuremath{\tan\beta}}
\newc{\azero}{\ensuremath{ A_0}}
\newc{\signmu}{\ensuremath{\rm{sgn}\,\mu}}
\newc{\atau}{\ensuremath{{A_{\tau}}}}
\newc{\mueff}{\ensuremath{\mu_{\rm{eff}}}}
\newc{\lam}{\ensuremath{{\lambda}}}
\newc{\kap}{\ensuremath{{\kappa}}}
\newc{\alam}{\ensuremath{{A_{\lambda}}}}
\newc{\akap}{\ensuremath{{A_{\kappa}}}}
\newc{\hs}{\ensuremath{ H_s}}      
\newc{\mhs}{\ensuremath{ m_{H_s}}} 
\newc{\mgut}{\ensuremath{ M_{\rm GUT}}}
\newc{\gut}{\ensuremath{{\rm GUT}}}
\newc{\mplanck}{\ensuremath{ M_{\rm P}}}      \newc{\mpl}{\ensuremath{ M_{\rm Pl}}}
\newc{\msusy}{\ensuremath{ M_{\rm SUSY}}}      \newc{\ms}{\ensuremath{ M_{\rm S}}}
 \newc{\hu}{\ensuremath{ H_u}}       \newc{\hd}{\ensuremath{ H_d}}
 \newc{\mhu}{\ensuremath{ m_{H_u}}}       \newc{\mhd}{\ensuremath{ m_{H_d}}}
 \newc{\mhuew}{\ensuremath{ m^{\ast}_{H_u}}}       \newc{\mhdew}{\ensuremath{ m^{\ast}_{H_d}}}
 \newc{\mhuewsq}{\ensuremath{ m^{\ast\, 2}_{H_u}}}       \newc{\mhdewsq}{\ensuremath{ m^{\ast\, 2}_{H_d}}}
 \newc{\mhl}{\ensuremath{m_\hl}} 
 \newc{\mhone}{\ensuremath{m_{h_1}}} 
 \newc{\mhtwo}{\ensuremath{m_{h_2}}} 
 \newc{\mhi}{\ensuremath{m_{\tilde{h}}}} 
 \newc{\mul}{\ensuremath{m_{\tilde{u}_L}}} 
 \newc{\mbone}{\ensuremath{m_{\tilde{b}_1}}}  
 \newc{\mtone}{\ensuremath{m_{\tilde{t}_1}}} 
 \newc{\ma}{\ensuremath{m_A}} 
 \newc{\mH}{\ensuremath{m_H}} 
 \newc{\maone}{\ensuremath{m_{a_1}}} 
 \newc{\matwo}{\ensuremath{m_{a_2}}}
 \newc{\hone}{\ensuremath{h_1}}
 \newc{\htwo}{\ensuremath{h_2}}
 \newc{\aone}{\ensuremath{a_1}}
 \newc{\atwo}{\ensuremath{a_2}}
 \newc{\mqthree}{\ensuremath{m_{\tilde{Q}_3}^2}}
 \newc{\muthree}{\ensuremath{m_{\tilde{u}_3}^2}}
 \newc{\mqli}{\ensuremath{m_{\tilde{Q}_{i}}}}
 \newc{\muri}{\ensuremath{m_{\tilde{u}_{i}}}}
 \newc{\mdri}{\ensuremath{m_{\tilde{d}_{i}}}}
 \newc{\mlli}{\ensuremath{m_{\tilde{L}_{i}}}}
 \newc{\meri}{\ensuremath{m_{\tilde{e}_{i}}}}
 \newc{\ts}{\ensuremath{T_{SUSY}}}

\newc{\sigsip}{\ensuremath{\sigma^{\rm SI}_{p}}}	\newc{\sigsin}{\ensuremath{\sigma^{\rm SI}_{n}}}
\newc{\sigsdp}{\ensuremath{\sigma^{\rm SD}_{p}}}	\newc{\sigsdn}{\ensuremath{\sigma^{\rm SD}_{n}}}
\newc{\sigsi}{\ensuremath{\sigma^{\rm SI}}}	\newc{\sigsd}{\ensuremath{\sigma^{\rm SD}}}
\newc{\abund}{\ensuremath{ \Omega h^2}}
\newc{\omegadm}{\ensuremath{ \Omega_{{\rm DM}}}}     \newc{\abunddm}{\ensuremath{ \Omega_{{\rm DM}} h^2}} 
\newc{\omegam}{\ensuremath{ \Omega_{{\rm m}}}}       \newc{\abundm}{\ensuremath{ \Omega_{{\rm m}} h^2}}
\newc{\omegab}{\ensuremath{ \Omega_{{\rm b}}}}	\newc{\abundb}{\ensuremath{ \Omega_{{\rm b}} h^2}}
\newc{\omegatot}{\ensuremath{ \Omega_{{\rm TOT}}}}
\newc{\omegacdm}{\ensuremath{ \Omega_{{\rm CDM}}}}   \newc{\abundcdm}{\ensuremath{ \Omega_{{\rm CDM}} h^2}}
\newc{\omegalambda}{\ensuremath{ \Omega_{\Lambda}}} \newc{\abundlambda}{\ensuremath{ \Omega_{\Lambda} h^2}}
\newc{\omegarad}{\ensuremath{ \Omega_{{\rm rad}}}}  \newc{\abundrad}{\ensuremath{ \Omega_{{\rm rad}} h^2}}
\newc{\rhocrit}{\ensuremath{ \rho_{\rm crit}}}
\newc{\rhochi}{\ensuremath{ \rho_{\chi}}}
\newc{\abunchi}{\ensuremath{\Omega_\chi h^2}}
\newc{\abundlsp}{\ensuremath{\Omega_{\rm LSP}h^2}}

\newc{\amu}{\ensuremath{ a_{\mu}}}        \newc{\amususy}{\ensuremath{ a_{\mu}^{\mathrm{SUSY}}}}
\newc{\amuexpt}{\ensuremath{ a_{\mu}^{\mathrm{expt}}}}        \newc{\amusm}{\ensuremath{ a_{\mu}^{\mathrm{SM}}}}
\newc\deltaamu{\ensuremath{\Delta a_{\mu}}} \newc{\deltaamususy}{\ensuremath{\delta a_{\mu}^{\mathrm{SUSY}}}}
\newc\gmtwo{\ensuremath{ (g-2)_{\mu}}} 
\newc{\deltagmtwomususy}{\ensuremath{\delta\left(g-2\right)_{\mu}^{\mathrm{SUSY}}}}
\newc{\deltagmtwomu}{\ensuremath{\delta\left(g-2\right)_{\mu}}}
\newc\BR{\ensuremath{\rm BR}}
\newc\bsgamma{\ensuremath{ b\rightarrow s \gamma }}
\newc\bxsgamma{\ensuremath{\overline{B}\rightarrow X_{s}\gamma}}
\newc\brbsgamma{\ensuremath{\BR\left(\bsgamma\right)}}
\newc\brbxsgamma{\ensuremath{\BR\left(\bxsgamma\right)}}
\newc\bsmumu{\ensuremath{B_s\to\mu^+\mu^-}}
\newc\brbsmumu{\ensuremath{\BR\left(B_s\to\mu^+\mu^-\right)}}
\newc\bdmmumu{\ensuremath{\overline{B}_d\to\mu^+\mu^-}}
\newc\bbbarmix{\ensuremath{\overline{B}_s\mbox{-}B_s}}      
\newc\delmbs{\ensuremath{\Delta M_{B_s}}}
\newc{\butaunu}{\ensuremath{B_u \rightarrow \tau \nu}}
\newc{\brbutaunu}{\ensuremath{\BR\left(B_u \rightarrow \tau \nu\right)}}

\newcommand*{\reftable}[1]{Table~\ref{#1}}         
       
\newcommand*{\reffig}[1]{Fig.~\ref{#1}}

\newcommand*{\refref}[1]{Ref.\cite{#1}}            
\newcommand*{\non}{\nonumber}

\newcommand*{\neutone}{\ensuremath{\tilde{\chi}^0_1}}
\newcommand*{\neuttwo}{\ensuremath{\tilde{{\chi}}^0_2}}

\newcommand*{\charone}{\ensuremath{\tilde{{\chi}}^{\pm}_1}}

\newcommand*{\thteen}{\ensuremath{\sqrt{s}=13\tev}}

\newcommand*{\pythia}{\text{PYTHIA}}

\newcommand*{\delphes}{\text{DELPHES 3}}

\let\oldcite\cite
\renewcommand*{\cite}{~\oldcite}

\newcommand*{\hl}{\ensuremath{h}}

\newcommand*{\met}{E_T^{\textrm{miss}}}
\newcommand*{\ptone}{p_T({\textrm{jet}_1})}
\newcommand*{\pttwo}{p_T({\textrm{jet}_2})}
\newcommand*{\mindfi}{(\Delta\phi(\met,\textrm{jet}))_{\textrm{min}}}
\newcommand*{\metht}{\met/\sqrt{H_T}}
\newcommand*{\metmef}{\met/\meff}
\newcommand*{\meff}{m_{\textrm{eff}}(\textrm{incl})}
\newcommand*{\meffj}{m^{\textrm{4j}}_{\textrm{eff}}}
\newcommand*{\dfij}{\Delta\phi^{\textrm{4j}}_{\textrm{min}}}
\newcommand*{\mtb}{m_{T,\textrm{min}}^b}
\newcommand*{\amt}{am_{T2}}
\newcommand*{\mtt}{m_{T2}}
\newcommand*{\mctop}{m_{\textrm{top}}^\chi}
\newcommand*{\delR}{\Delta R(b,l)}
\newcommand*{\dfi}{\Delta\phi(\met,\textrm{jet})}

\begin{document}

\begin{flushright}
DO-TH 16/18 \\
QFET-2016-15
\end{flushright}

\title{Phenomenological MSSM in light of new 13 TeV LHC data}  
\author{Kamila Kowalska\\[2ex]
\small\it Fakult\"at f\"ur Physik, TU Dortmund, Otto-Hahn-Str.4, D-44221 Dortmund, Germany\\
}
\date{}

{\let\newpage\relax\maketitle}
\centering
\url{kamila.kowalska@tu-dortmund.de}

\abstract{

We present the first analysis of the p19MSSM with neutralino dark matter, in light of 13 TeV LHC data with an integrated luminosity 
of $\sim14$\invfb. We recast twelve experimental analyses performed by the ATLAS collaboration and derive exclusion bounds on the parameter space of the model. We find that 25\% of the model points
can be excluded at  95\% \cl\ by a combination of the implemented searches. The spectrum allowed after the 13 TeV results are taken into account presents 
stops heavier than at least 400\gev, gluinos heavier than at least 790\gev, and squarks of the first and second generation heavier than at least 440\gev.

\section{Introduction}\label{intro}

In 2015 the LHC entered the second phase of its operation. The center-of-mass energy of the proton-proton collisions has increased to \thteen.
By now, both ATLAS and CMS collaborations have collected a large amount of data corresponding to an integrated luminosity of around 14\invfb. 
The experimental analyses aimed at interpreting these results in the framework of beyond-the-Standard Model (BSM) scenarios, including supersymmetry (SUSY), 
extra dimensions, leptoquarks, and many others. Albeit some of the searches present mild upward fluctuations in the number of signal events over the background yields, 
none of them has reached a statistical significance that would allow to claim, or at least to suspect, a discovery of a new particle.



In this paper we study the impact of the new 13 \tev\ LHC data with luminosity of $\sim$14\invfb\ on the phenomenological Minimal Supersymmetric 
Standard Model (pMSSM)\cite{Berger:2008cq} characterized by 19 independent parameters.
To this end, we use a sample of model points generated by the ATLAS collaboration in Ref.\cite{Aad:2015baa} that satisfies a set of experimental constraints from dark matter
searches, Higss measurements, electroweak and flavour physics, as well as includes the exclusion bounds from the 8\tev\ LHC SUSY searches.
The previous LHC study of this dataset with \thteen\ and luminosity of 3.2\invfb\ has been performed in Ref.\cite{Barr:2016inz}.

The model-dependent limits from the experimental SUSY searches are interpreted by the collaboratios in the so-called Simplifed Model scenarios (SMS)\cite{Chatrchyan:2013sza}, which make simplifying assumptions about the masses of SUSY particles.
In most scenarios the SMS spectrum consists of only several light states, one of them being the lightest neutralino.
If, however, also other light particles are present, the derived exclusion limits can be altered. 
Therefore, in order to evaluate in a complete and self-consistent way the impact of the LHC SUSY searches on the parameter space of p19MSSM,
we reinterpret the SMS results by simulating in detail the experimental searches with a likelihood function approach\cite{Fowlie:2012im,Fowlie:2013oua,Kowalska:2013ica}.


This paper is organized as follows. In Section \ref{num} we describe the model sample used in the study and make a brief overview of the numerical tools and procedures used  to implement the 
LHC SUSY searches. Section \ref{res} is dedicated to a discussion of our main results. We summarize our findings in Section \ref{sum}.

\section{Methodology}\label{num}
\subsection{p19MSSM ATLAS sample} 

The phenomenological MSSM (pMSSM) is characterized by 19 independent SUSY parameters defined at the scale \msusy, which is the geometrical
average of the physical stop masses. The number of free parameters results from
 the following assumptions: $R$-parity is conserved, all soft parameters are real, the off-diagonal elements of soft matrices are set to zero in order
to ensure Minimal Flavor Violation, neutralino is the lightest supersymmetric particle (LSP), and two first generations of squarks and leptons are degenerate
in order to suppress potentially large SUSY contributions to the Flavor Changing Neutral Currents processes. 

In this study we use a sample of p19MSSM model points provided by the ATLAS collaboration in Ref.\cite{Aad:2015baa} and generated using methods similar to those
described in\cite{Berger:2008cq,CahillRowley:2012cb,CahillRowley:2012kx,Cahill-Rowley:2014twa}. All masses and trilinear couplings where
scanned up to 4 TeV, which results in the range of superpartner masses that falls within the reach of the LHC. 
The only exception was the trilinear coupling $A_t$, scanned up to 8 TeV in order to boost the Higgs boson mass to 125 GeV through the stop loops.
The lower bounds imposed on the soft masses came from the direct SUSY searches at LEP\cite{LEP}.
The Standard Model parameters where fixed at their central values given in Ref.\cite{Agashe:2014kda}.

Additional phenomenological constraints from the dark matter (DM) searches (relic density, spin independent nucleon cross section, 
spin dependent proton cross sections), 
flavor physics (\brbsgamma, \brbsmumu, \brbutaunu, \deltagmtwomususy) and the electroweak precision data ($\Delta\rho$) were also taken into account, 
with the acceptance ranges given in \reftable{tab:ATLAS_exp}. 
Finally, the impact of 22 ATLAS SUSY searches at $\sqrt{s}=7$ and $8$ \tev\ with integrated luminosity of  20.3\invfb\
was evaluated. 
The resulting p19MSSM sample contains 183,030 allowed model points.
33.5\% of them is characterized by bino-like LSP, 42.6\% by higgsino-like LSP, and 23.9\% are the points for which the 
neutralino LSP is wino. 
Note that the type of neutralino LSP has been defined by ATLAS according to
the maximal neutralino mixing matrix parameter. As a consequence, not all the points in a given type sample are pure gaugino eigenstates.

\begin{table}[b]\footnotesize
\begin{centering}
\begin{tabular}{|c|c|c|c|}
\hline 
Constraint & Min & Max & Ref. (exp., theo.)\\
\hline
\abund & - & 0.1208 & \cite{Ade:2015xua} \\
\sigsip, \sigsdp & - & See Ref.\cite{Aad:2015baa} & \cite{Akerib:2013tjd},\cite{Behnke:2012ys,Aprile:2013doa}\\
$\Delta\rho$ & -0.0005 & 0.0017 & \cite{Baak:2012kk}\\
\brbsgamma & $2.69\times 10^{-4}$ & $3.87\times 10^{-4}$ & \cite{Amhis:2012bh}\\
\brbsmumu & $1.6\times 10^{-9}$ & $4.2\times 10^{-9}$ & \cite{DeBruyn:2012wk,CMS:2014xfa} \\
\brbutaunu & $6.6\times 10^{-5}$ & $16.1\times 10^{-5}$ & \cite{Charles:2004jd,Aubert:2009wt,Hara:2010dk,Lees:2012ju,Adachi:2012mm}\\
\deltagmtwomususy & $-17.7\times 10^{-10}$ & $43.8\times 10^{-10}$ & \cite{Czarnecki:2002nt,Bennett:2004pv,Bennett:2006fi,Nyffeler:2009tw,
Hagiwara:2011af,Aoyama:2012wk}\\
$m_h$ & 124\gev\ & 128\gev\ & \cite{Heinemeyer:2011aa,Aad:2015zhl} \\
\hline
\end{tabular}
\caption{\footnotesize The lower and upper bounds on the relevant phenomenological observables as used in Ref.\cite{Aad:2015baa}. }
\label{tab:ATLAS_exp}
\end{centering}
\end{table}

The lightest squarks and gluinos allowed by the 8 \tev\ data correspond to the compressed spectra region which is notoriously difficult 
to probe at the LHC. The reason is a small mass difference between the color sparticles and 
the lightest neutralino that results in very soft decay products. In fact, if this mass difference is very small, the next-to-LSP particle becomes 
semi-stable and can not be tested by the standard SUSY searches. For this reason, following the 
approach of\cite{Aad:2015baa}, we remove 3,500 of such models from the sample.
After that, the minimal masses of squarks and gluinos found in the p19MSSM set, together with the corresponding LSP mass, read: 
\bea\label{masses_8}
\mtone^{\textrm{min}}=275\gev,\;\;m_{\neutone}=243\gev,\non\\
\mbone^{\textrm{min}}=234\gev,\;\;m_{\neutone}=220\gev,\non\\
m_{\tilde{d}_R}^{\textrm{min}}=291\gev,\;\;m_{\neutone}=272\gev,\non\\
m_{\tilde{u}_R}^{\textrm{min}}=251\gev,\;\;m_{\neutone}=222\gev,\non\\
m_{\tilde{u,d}_L}^{\textrm{min}}=302\gev,\;\;m_{\neutone}=266\gev,\non\\
\mgluino^{\textrm{min}}=566\gev,\;\;m_{\neutone}=513\gev.
\eea
On the other hand, in the region of the parameter space where neutralino LSP is light ($m_{\neutone}\leq 50\gev)$ and sensitivity of the LHC SUSY searches reaches its maximum, 
the limits become significantly stronger and the minimal masses read:
\be
\mtone^{0}=553\gev,\;\; \mbone^{0}=498\gev,\;\; m_{\tilde{d}_R}^{0}=633\gev,\;\; m_{\tilde{u}_R}^{0}=739\gev,\;\;\mgluino^{0}=1188\gev.
\ee

\subsection{13 TeV LHC SUSY searches}\label{lhc}

Both ATLAS and CMS collaborations performed a large number of analyses at \thteen\ that cover a wide spectrum of experimental signatures. 
For the purpose of this paper, we decided to implement the ATLAS searches only, in order to be consistent with the results of the 8 TeV analyses 
incorporated in the p19MSSM sample described in the previous subsection. 

From eleven 3.2\invfb\ ATLAS searches within the R-parity conserving MSSM for which the experimental analysis are available, we implemented seven
which exhibit both the strongest exclusion limits for a given production scenario and the largest expected sensitivity in beyond the SMS framework. 
We used here as a helpful guideline the results of 
Ref.\cite{Barr:2016inz} where the impact of six ATLAS searches on the same p19MSSM sample has been quantified. 
Five of the above analysis have been recently upated with 13-14 \invfb\ of data. Additionally, one 13.3 \invfb\ study appeared that does not have the lower luminosity equivalent.
Below we present the full list of the implemented analyses:
\begin{itemize}
\item ATLAS 0 leptons + 2-6 jets + $\met$, 3.2\invfb\cite{Aaboud:2016zdn}, 13.3\invfb\cite{ATLAS-CONF-2016-078},
\item ATLAS 1 lepton + jets + $\met$, 3.2\invfb\cite{Aad:2016qqk}, 14.8\invfb\cite{ATLAS-CONF-2016-054},
\item ATLAS 3 b-tagged jets + $\met$, 3.2\invfb\cite{Aad:2016eki}, 14.8\invfb\cite{ATLAS-CONF-2016-052},
\item ATLAS 0 lepton + (b)jets + $\met$, 13.3\invfb\cite{ATLAS-CONF-2016-077},
\item ATLAS 1 lepton + (b)jets + $\met$, 3.2\invfb\cite{Aaboud:2016lwz}, 13.3\invfb\cite{ATLAS-CONF-2016-050},
\item ATLAS 2 leptons + jets + $\met$, 3.2\invfb\cite{ATLAS-CONF-2016-009}, 
\item ATLAS 2 b-tagged jets + $\met$, 3.2\invfb\cite{Aaboud:2016nwl},
\item ATLAS monojet + $\met$, 3.2\invfb\cite{Aaboud:2016tnv}.
\end{itemize}

New results from the LHC 13 TeV are taken into account by employing the recast procedure developed and described in detail in 
Refs.\cite{Fowlie:2012im,Fowlie:2013oua,Kowalska:2013ica}. Its key element is a construction of an approximate but accurate likelihood function,
which yields an exclusion \cl\ (confidence level) for each point in the analyzed p19MSSM sample. 

In order to obtain the likelihood function, one needs to mimic the analyses performed by the experimental collaborations. 
For every point in the parameter space the spectrum is generated with \texttt{softsusy}\cite{Allanach:2001kg} and the decay branching ratios calculated with
\texttt{SUSYHIT}\cite{Djouadi:2006bz} (alternatively, the provided SLHA spectrum file is used). Next, $2\times10^4$ events are generated at the parton level with \texttt{\pythia8}\cite{Sjostrand:2007gs} and the
hadronization products are passed to the fast detector simulator \texttt{\delphes}\cite{deFavereau:2013fsa} to reconstruct the physical objects. 
The ATLAS detector card is used, with the settings adjusted to those recommended by the experimental collaboration.
The $b$-tagging algorithm used by \texttt{\delphes} is tuned to match the corresponding efficiencies reported by ATLAS for the 13 TeV data\cite{ATL-PHYS-PUB-2015-022}. 

The physical objects produced by the detector simulator are used to construct a set of kinematical variables
proper of a given SUSY search and 
chosen to reduce the SM background by applying a series of the selection cuts. 
Finally, different kinematical bins $i$ are defined and the efficiencies $\varepsilon_i$ are evaluated 
as the fraction of events that passes all the cuts. 

The number of signal events in a given bin is determined as $s_i=\varepsilon_i\times\sigma_{\textrm{LO}}\times\int L$, where $\int L$ 
is the integrated luminosity and $\sigma_{\textrm{LO}}$ the leading order (LO) cross-sections.
We decided to use the LO cross-section instead of the next-to-LO (NLO) one as a good compromise between the accuracy of the results and the fact how time-consuming their derivation is.
The SUSY spectrum that arises in the framework of the p19MSSM is in general complex and different particles contribute to the total production cross-section. 
In order to incorporate the NLO corrections in a consistent way, one should either decompose each model into various production channels and use the official 
values of the NLO+NLL cross-sections provided by  LHC SUSY Cross Section Working Group\cite{LHCSXSECWG}, or interface the whole package with a numerical tool that calculate the 
NLO cross-section directly. 
On the other hand, we have checked in the framework of several SMS that when the LO cross-section is used instead of the NLO one, the resulting exclusion bounds on the sparticle mass are weakened
by around 50-100 GeV in the part of the parameter space where neutralino is light. Since the recast procedure employed in this study is designed as an approximation of the experimental analyses, 
we treat underestimation of the total cross-section as an additional source of uncertainty and interpret our exclusion limits as conservative ones.

The signal is statistically compared to the numbers of the observed ($o_i$) and background ($b_i$) events, given in the experimental papers, 
through the Poisson distribution $P$. 
The systematic uncertainties on the background yields ($\delta b_i$) are taken into account
by convolving $P$ with the Gaussian distribution $G$.
The likelihood function for each bin is thus calculated as,
\begin{equation}
\mathcal{L}_i(o_i, s_i, b_i)=\int P(o_i|s_i,\bar{b}_i)G(\bar{b}_i|b_i,\delta b_i)d\bar{b}_i\,.\label{likelihood}
\end{equation}
The total likelihood function ${\cal L_{\textrm{LHC}}}$ for a given model point is calculated either as a product of the likelihoods from each 
signal bin, or as the likelihood from the bin with the best expected sensitivity, depending on whether the bins are exclusive or inclusive.
The appropriate exclusion \cl\ is obtained from the $\delta\chisq$ variable as $\delta\chisq=-2\log(\mathcal{L_{\textrm{LHC}}}/\mathcal{L}_{0})$, where
$\mathcal{L}_{0}$ corresponds to the background-only hypothesis.

Note that when the observed number of events is large enough, this approach is bound to give
exactly the same results as the CL$_s$ method\cite{0954-3899-28-10-313} employed by the experimental collaborations
in order to derive the C.L. intervals. This is a direct consequence of the central limit theorem, which
roughly states that if the data is independent its distribution will converge to the normal distribution
in the limit of high statistics. The validity of this simplification should be verified case by
case with the proper validation of the results. In practice, however, at the level of precision expected for analyses like this, no significant discrepancies 
between the two approaches can be registered for the number of the observed and background events reported in the SUSY searches.

A brief description of each implemented ATLAS search, as well as  validations of the recast procedure against the experimental results,
are given in Appendix \ref{13searches}.
\section{Results for SUSY spectra}\label{res}

In this section we analyze the impact of the implemented ATLAS searches on the p19MSSM parameter space 
allowed after inclusion of the 8 \tev\  run data. 
Since no statistically significant excess above the SM background expectation is observed, we derive the exclusion limits only.

\begin{table}[t]\footnotesize
\begin{centering}
\begin{tabular}{|c|c|c|c|c|}
\hline 
 & Bino & Higgsino & Wino & ALL \\
 \hline
Search  & \multicolumn{4}{|c|}{3.2\invfb} \\
\hline
0 leptons + 2-6 jets + $\met$ & 17.9\%	& 10.3\% & 8.0\% & 12.3\% \\
1 lepton + jets + $\met$ & 0.9\% & 1.0\% & 0.4\% & 0.8\% \\ 
3 b-tagged jets + $\met$ & 2.4\% & 3.2\% & 1.0\% & 2.4\%\\
1 lepton + (b)jets + $\met$ & 0.4\% & 0.3\% & 0.1\% & 0.3\% \\
2 leptons + jets + $\met$ & 0.1\% & 0.2\% & 0.1\% & 0.1\% \\
2 b-tagged jets + $\met$ & 1.3\% & 0.3\% & 0.3\% & 0.6\%\\
monojet + $\met$ &  2.6\% & 0.1\% & 0.3\%  & 1.0\%\\
\hline
\bf{Total  3.2\invfb} & \bf{19.8}\%	& \bf{12.3}\% &  \bf{8.6}\% & \bf{14.5}\%\\
\hline
 & \multicolumn{4}{|c|}{$\sim$14\invfb}\\
\hline
0 leptons + 2-6 jets + $\met$ & 30.6\%	& 15.6\% & 10.3\% & 18.6\%\\
1 lepton + jets + $\met$ & 7.2\% & 11.3\% & 2.9\% & 7.7\%\\ 
3 b-tagged jets + $\met$  & 3.4\% & 5.0\% & 1.4\% & 3.5\%\\
1 lepton + (b)jets + $\met$ & 3.1\% & 6.0\% & 1.2\% & 3.8\%\\
0 lepton + (b)jets + $\met$ & 5.6\% & 4.5\% & 1.4\% & 4.1\%\\
\hline
\bf{Total  $\sim$14\invfb} & \bf{34.8}\%	& \bf{25.5}\% &  \bf{13.5}\% & \bf{24.7}\%\\
\hline
\end{tabular}
\caption{\footnotesize Percentage of the p19MSSM points that are excluded by 13 TeV searches at  95\% \cl. The  results for bino-, higgsino- and wino-like neutralino LSP, as well as for the whole p19MSSM sample, are presented in the consecutive columns.  }
\label{tab:ATLAS_LHC_13TeV}
\end{centering}
\end{table}

In \reftable{tab:ATLAS_LHC_13TeV} we show the percentage of model points that have been excluded at 95\%  \cl\ by 
each of the implemented searches. To capture a dependence of the exclusion bounds on the properties of the neutralino LSP, we present the results for the bino-, higgsino- and wino-like LSP cases separately. In the last column of \reftable{tab:ATLAS_LHC_13TeV} the results for the whole p19MSSM sample are indicated.
To quantify the total impact of the LHC 13 TeV data, we combine the individual searches through ``the-best-of'' strategy, using for each model point the result from the search that presents the best sensitivity. 

%

Only 14.5\% of the p19MSSM models can be excluded at 95\% \cl\ by the LHC SUSY searches with 3.2\invfb\ of the 13 TeV data. The percentage increases to 19.8\% if only
bino-like LSP is taken into account, while for the higgsino-like LSP it is reduced to 12.3\%. Those numbers are consistent with findings of \refref{Barr:2016inz} and we will make a more detailed comparison of the two sets of results later on. A difference in sensitivity of the LHC searches for different types of the LSP comes from the fact that the number of models with gluinos lighter than 1 TeV is much higher in the bino-like case than in the higgsino-like one, as already noted in \refref{Barr:2016inz}. That is due to the fact that if the LSP is bino-like, an efficient mechanism of co-annihilation must exist that allows to reduce the relic density of the dark matter below the critical value, to avoid overclosing the universe. Indeed, almost all light gluinos find in the p19MSSM sample are very close in mass to the neutralino LSP.

In the case of wino-like neutralino LSP, we find out that only 8.6\% of the parameter space is excluded with the luminosity of 3.2\invfb. That result is by a 
factor of 1.7 lower than the corresponding number quoted in \refref{Barr:2016inz}. Since such a reduction in sensitivity is observed for all the searches
that were also considered in \refref{Barr:2016inz}, a discrepency most likely arises at the level of events generation\footnote{\refref{Barr:2016inz} uses 
MadGraph5 to generate events and \pythia\ 6.428 for showering and hadronization.}.

The most effective in setting the exclusion bounds are inclusive all-hadronic searches, both with b-tagged and non-b-tagged jets.
The former shows perfect agreement with the results of \refref{Barr:2016inz} (except when the LSP is wino-like). 
In the case of the latter, the percentage of excluded models calculated by us is lower than the ones of \refref{Barr:2016inz} by around 25\%. This 
discrepancy may result from a different tuning of the b-tagging algorithm used by \delphes.

Contrarily, with 3.2\invfb\ of data the 1 lepton searches can test only a small fraction of the parameter space, as in the p19MSSM the branching ratio for
the color particles decaying to chargino (that would lead to a leptonicaly  decaying $W$ boson) is much smaller than the one directly to the LSP. 
Once more, our results are in agreement with the findings of \refref{Barr:2016inz}.

The monojet analysis tests the region of the compressed spectra, mainly through the production of the first- and second-generation squarks. 
The efficiency of this search in our study is slightly lower than the one observed in \refref{Barr:2016inz}, although one needs to keep in mind that this kind of analysis, 
being based on the initial state radiation jet identification, is particularly sensitive to the detector simulator settings. 

 \begin{figure}[t]
 \centering
 \subfloat[]{
 \label{fig:a_res1}
 \includegraphics[width=0.48\textwidth]{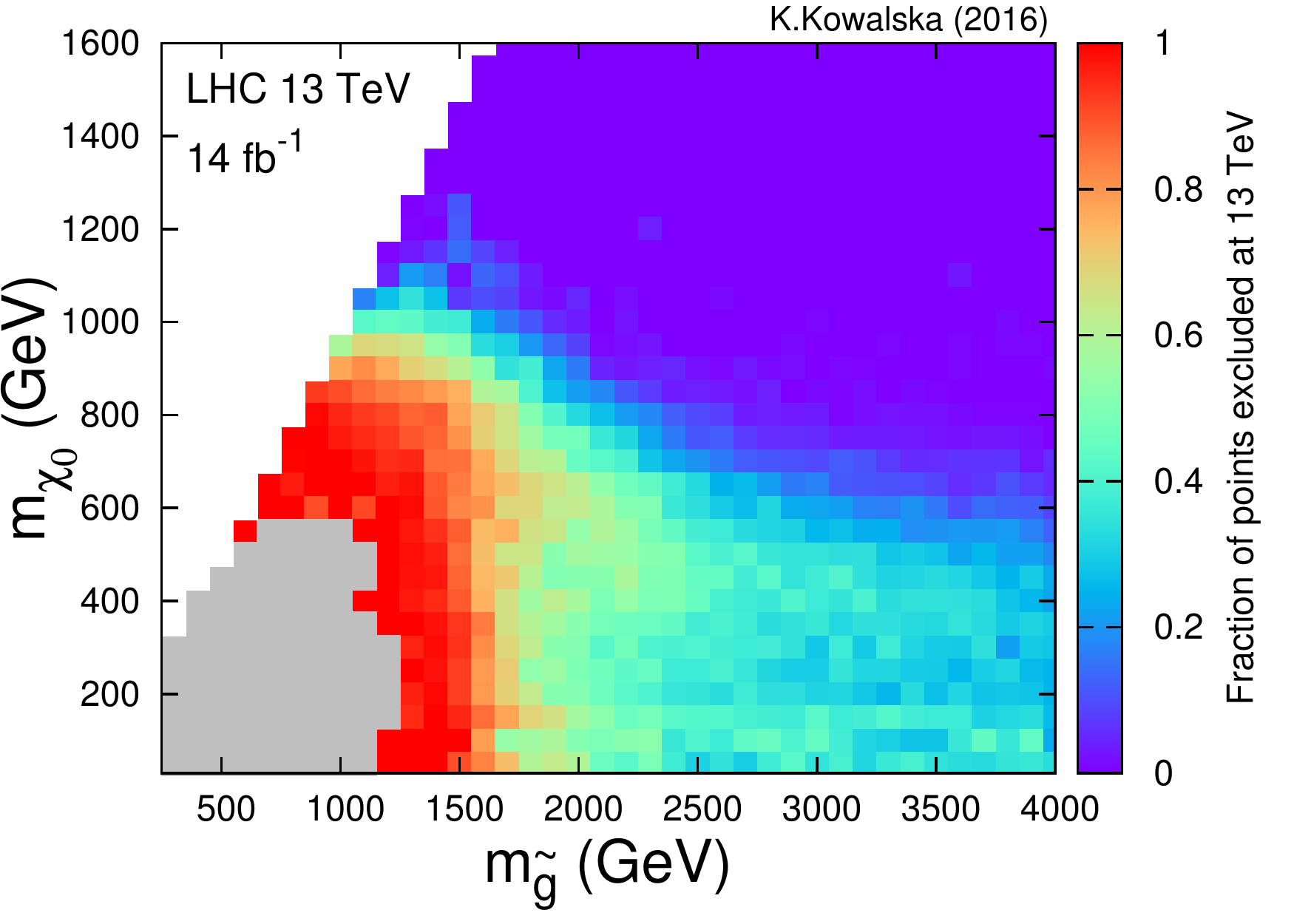}
 }
 \subfloat[]{
 \label{fig:b_res1}
 \includegraphics[width=0.48\textwidth]{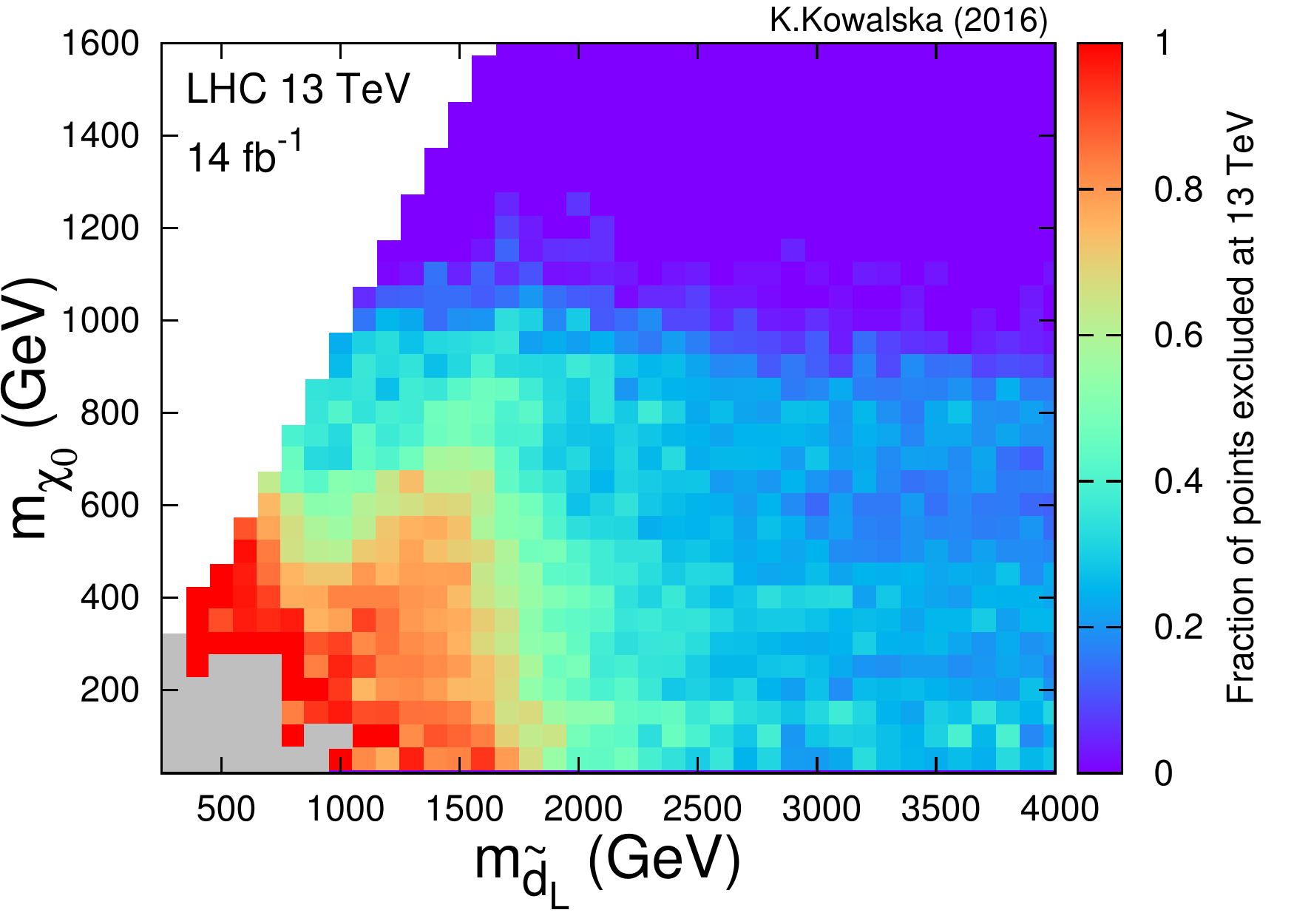}
 }\\
 \subfloat[]{
 \label{fig:c_res1}
 \includegraphics[width=0.48\textwidth]{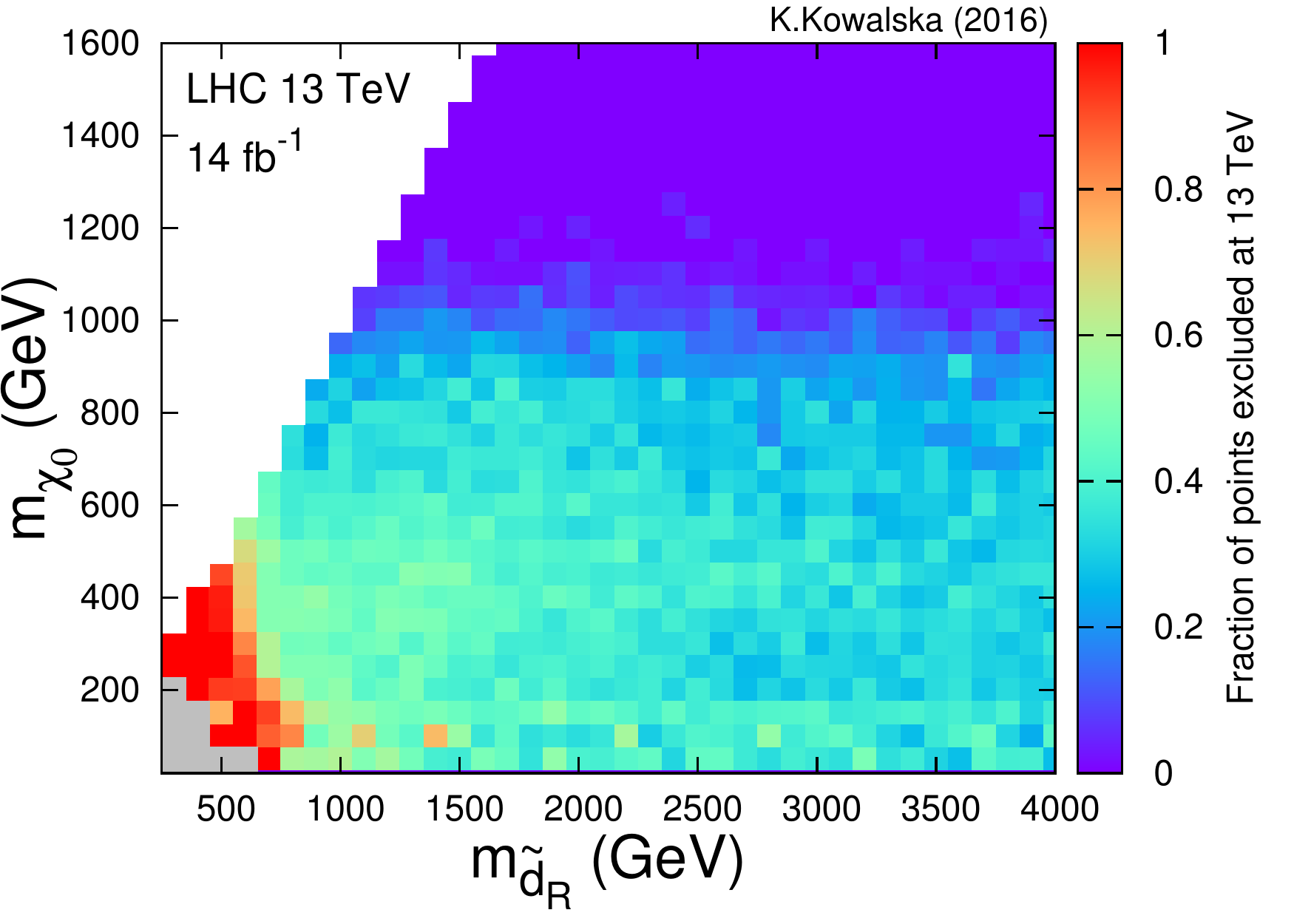}
 }
 \subfloat[]{
 \label{fig:d_res1}
 \includegraphics[width=0.48\textwidth]{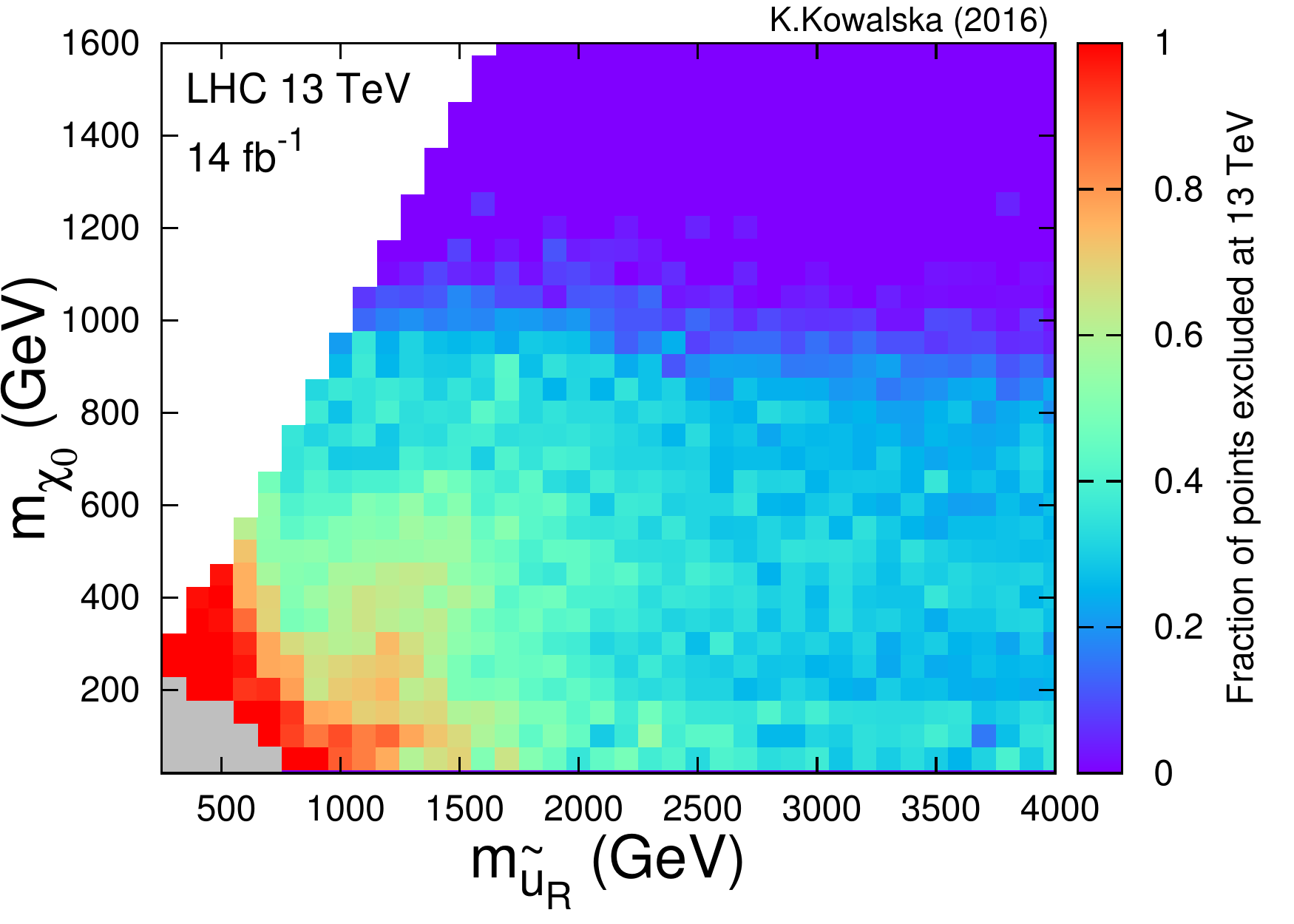}
 }
 \caption[]{\footnotesize Percentage of p19MSSM points excluded at 95\%\cl\ by a combination of twelve 13 TeV ATLAS SUSY searches projected into \subref{fig:a_res1} $(\mgluino,m_{\neutone})$ plane, \subref{fig:b_res1} $(m_{\tilde{d}_L},m_{\neutone})$  plane,  \subref{fig:c_res1} $(m_{\tilde{d}_R},m_{\neutone})$ plane, and \subref{fig:d_res1} $(m_{\tilde{u}_R},m_{\neutone})$ plane.
 While combining the individual searches, ``the-best-of'' strategy is employed that 
 uses for each model point the result from the search with the best sensitivity. Red color indicates those parts of the parameter space where 100\% of points is excluded. 
 Gray color corresponds to masses that have been excluded by a combination of 22 ATLAS SUSY searches at 8 TeV. }
 \label{fig:mglu_mchi_13TeV}
 \end{figure}

Finally, in the case of the 2 lepton search its  sensitivity over the p19MSSM parameter space is minimal. For this reason we do not include it in the higher luminosity update.

The sensitivity of the LHC SUSY searches based on the $\sim$14\invfb\ data sample is much stronger. After their inclusion 25.0\% of the p19MSSM model points are excluded at 95\% \cl. For the bino-like LSP the corresponding number increases to almost 35\%, while for the higgsino- and wino-like LSP 25.5\% and 13.5\% of the parameter space is excluded, respectively.

The improvement is particularly visible in the case of the 1 lepton + jets search, which is also confirmed by the ALTAS limits derived in the SMS framework. Such a significant increase of its reach results from the fact that a $\sim 2\sigma$ excess, observed in the 3.2\invfb\ data in one of the kinematical bins proper to this search, is now gone.
Contrarily, new all-hadronic analyses, in particular the one looking for 3-b jets signature, although exclude more models than their 3.2\invfb\ counterparts, 
gained less in terms of sensitivity due to the presence of several $\sim 2\sigma$ excesses that weaken the observed exclusions limits with respect to the expected ones. 

To get some qualitative feeling how the new LHC results affect the allowed mass ranges of the SUSY particles, we present their impact in two-dimensional projections of the full p19MSSM parameter space, following the style adopted in \refref{Aad:2015baa}. 
For each projection we bin the relevant sparticle masses, and for each bin we calculate the fraction of excluded models, defined as a ratio of the number of points excluded to the total number of points in a bin. 

In \reffig{fig:mglu_mchi_13TeV} we show the percentage of p19MSSM models that are excluded at 95\% \cl\ by ``the-best-of'' combination of the 13 TeV SUSY searches with 3.2\invfb\ and $\sim$14 \invfb\ of data in \subref{fig:a_res1} $(\mgluino,m_{\neutone})$ plane, in \subref{fig:b_res1} $(m_{\tilde{d}_L},m_{\neutone})$ plane, in \subref{fig:c_res1} $(m_{\tilde{d}_R},m_{\neutone})$ plane, and in \subref{fig:d_res1} $(m_{\tilde{u}_R},m_{\neutone})$ plane. In gray those mass bins are shown that has been excluded already at 8 TeV.
 Red color indicates those parts of the parameter space where 100\% of points is excluded and therefore can be interpreted as a 95\%\cl\ lower bound on the sparticle mass. 
 
One observes that the strongest and most solid exclusion limits can be derived in the case of the gluino mass, which reads $\mgluino\gsim 1.45\tev$ if neutralino is 
light ($m_{\neutone}\leq 50$ \gev), and $\mgluino\gsim 0.8\tev$ in the whole p19MSSM sample. 
The 95\% \cl\ exclusion bound coincide with the one derived in the framework of the SMS of Ref.\cite{ATLAS-CONF-2016-078} in the heavy neutralino region, while is weaker by around 300 GeV in the light neutralino region.
This is due to the fact that inclusive all hadronic searches, being sensitive to the number and energy of the outgoing jets only, can test a wide variety of possible decay chains and produce similar results both in the framework of the SMS and within the more general SUSY scenarios. 

In the case of the squarks of the first and second generations the exclusion limits are notably weaker than in the corresponding SMS of Ref.\cite{ATLAS-CONF-2016-078}.
The main reason is that all 8 suqarks are assumed to be degenerated in the SMS, while in the p19MSSM various mass hierarchies between the left- and right-handed superparters are 
possible. \reffig{fig:mglu_mchi_13TeV} emphasizes also the difference in sensitivity of the SUSY searches when left and right squarks are considered. The reason is 
well known and was explained, for instance, in Ref.\cite{Aad:2015baa}. Since the left-handed squarks are close in mass their production cross section is always
enhanced w.r.t. the right-handed ones. Moreover, since there are fewer valence down quarks than up in the proton, the limit of \reffig{fig:mglu_mchi_13TeV}
\subref{fig:c_res1} is slightly weaker than the limit of \reffig{fig:mglu_mchi_13TeV}\subref{fig:d_res1}.

In \reffig{fig:msq_mchi_13TeV} we show the percentage of p19MSSM models that are excluded at 95\% \cl\ in \subref{fig:a_res2} $(\mtone,m_{\neutone})$  and 
\subref{fig:b_res1} $(\mbone,m_{\neutone})$ planes.
Contrarily to the case of gluino and squarks of the first and second generations,  only a slight improvement  with respect to the 8 TeV results is observed in the limits on the stop mass, mainly in the part of the parameter space were neutralino is relatively heavy. Dedicated 0- and 1-lepton searches, while effective in the framework of the SMS with right stops and neutralinos only, lose sensitivity if a larger number of light particles is present, or if different amounts of mixing between two stops are considered.
 On the other hand, the compressed spectra region is well tested  by the monojet analysis. 

The 13 TeV limit on the sbottom mass is by around 50 \gev\ stronger with respect to the 8 TeV results in the part of the parameter space where neutralino is lighter than 300 \gev. The improvement reaches 100 GeV in the compressed spectra region. Sbottoms are predominantly excluded by a combination of a dedicated 2 b-jets search with 3.2\invfb\ of data, and the all-hadronic multijet search with 13.3\invfb, as both of them effectively test decay topologies characterized by the presence of two jets. When the former analysis is updated to higher luminosity, the exclusion limits should become even stronger.

 \begin{figure}[t]
 \centering
 \subfloat[]{
 \label{fig:a_res2}
 \includegraphics[width=0.48\textwidth]{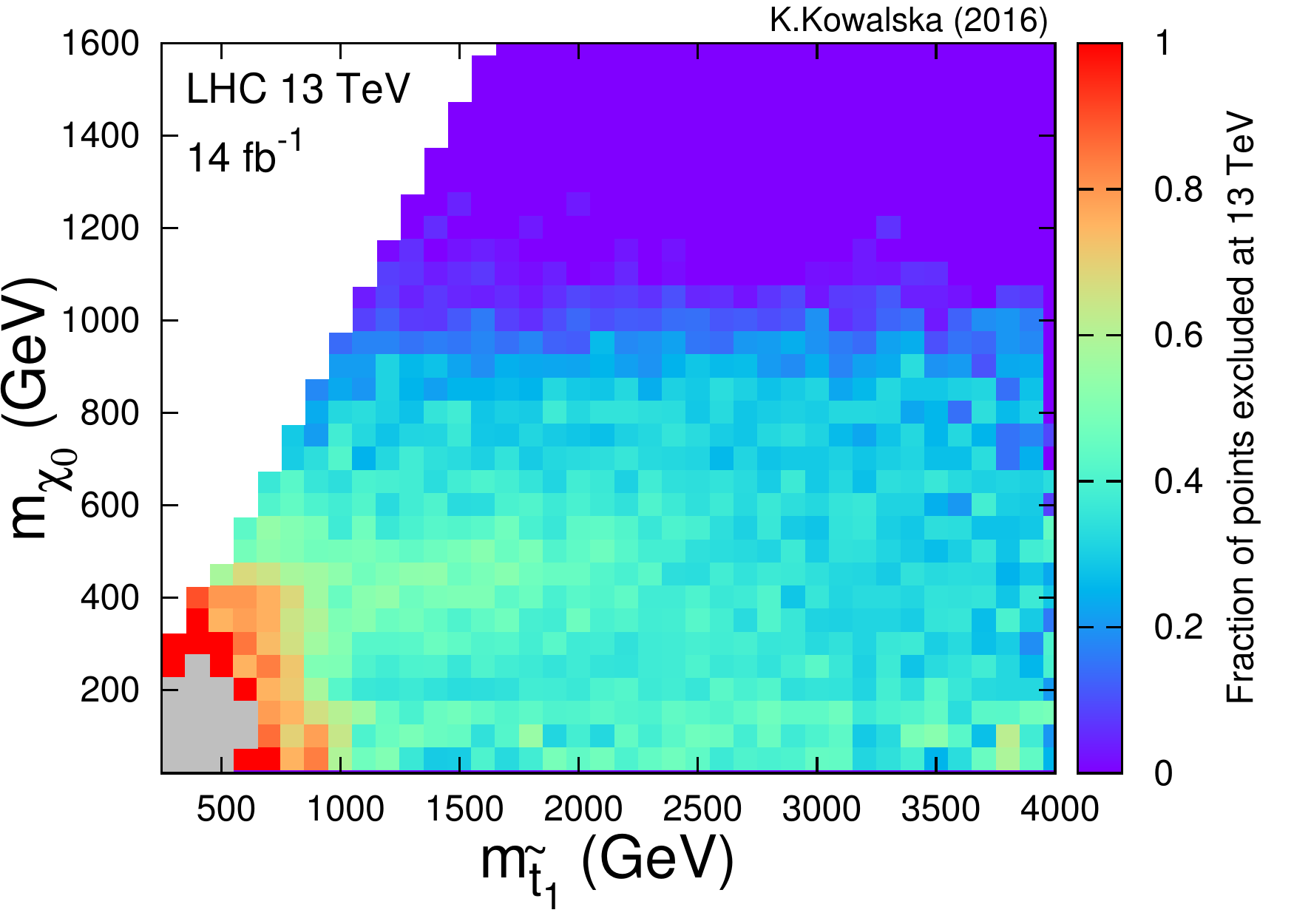}
 }
 \subfloat[]{
 \label{fig:b_res2}
 \includegraphics[width=0.48\textwidth]{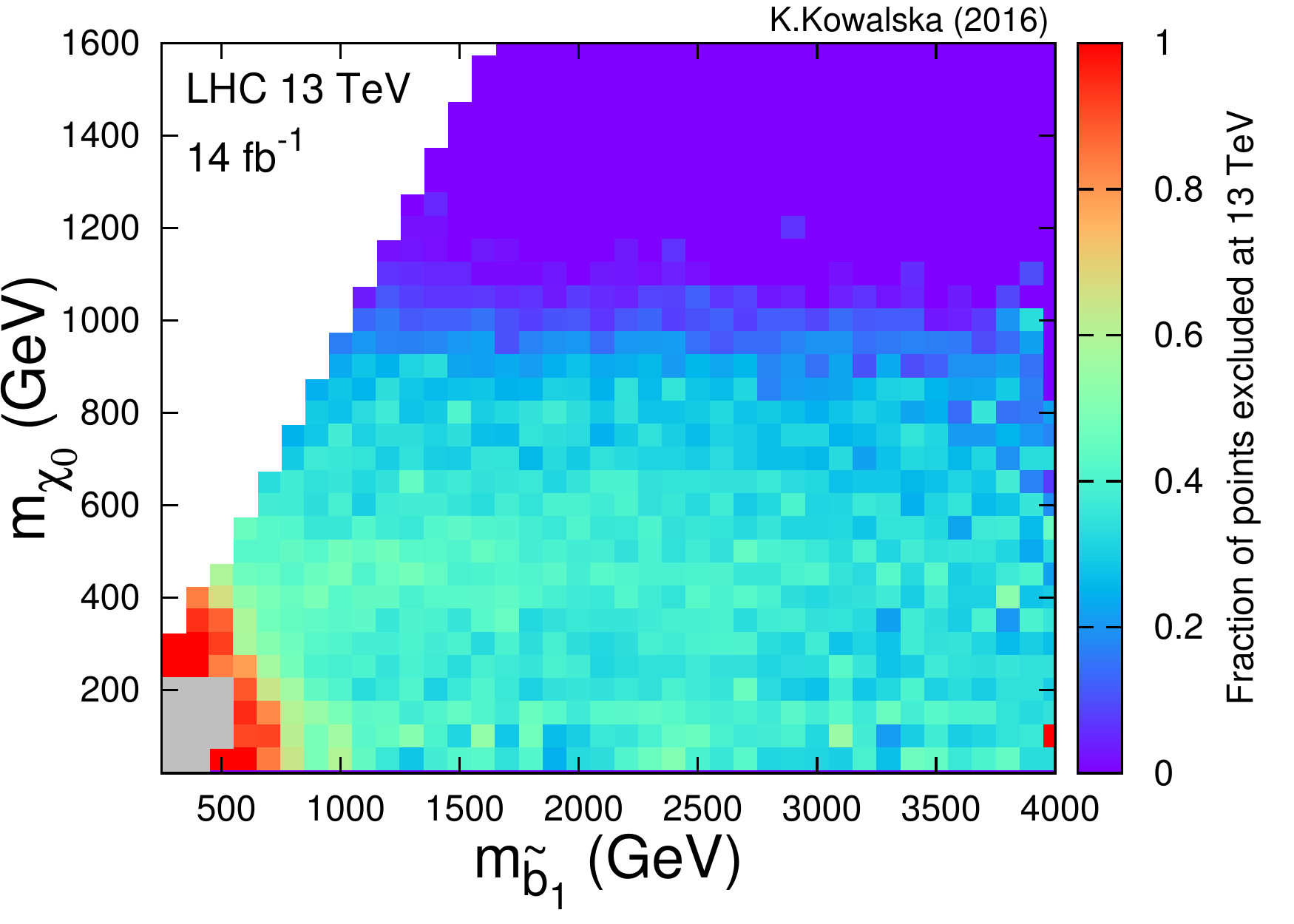}
 }
 \caption[]{\footnotesize Percentage of p19MSSM points excluded at 95\%\cl\ by a combination of twelve 13 TeV ATLAS SUSY searches projected into \subref{fig:a_res1} $(\mtone,m_{\neutone})$ plane, and \subref{fig:b_res1}  $(\mbone,m_{\neutone})$ plane.
  The color code is the same as in \reffig{fig:mglu_mchi_13TeV}. }
 \label{fig:msq_mchi_13TeV}
 \end{figure}

We conclude this section with quoting the minimal values of superparticle masses left in the p19MSSM sample after the 95\% \cl\ exclusion bounds from the LHC 13 TeV have been imposed:
\bea\label{masses_13}
\mtone^{\textrm{min}}=396\gev,\;\;m_{\neutone}=376\gev,\qquad\qquad \textrm{light neutralino:}\;&\mtone^{0}=744\gev,\non\\
\mbone^{\textrm{min}}=357\gev,\;\;m_{\neutone}=342\gev,\qquad\qquad \textrm{light neutralino:}\;&\mbone^{0}=648\gev,\non\\
m_{\tilde{d}_L}^{\textrm{min}}=483\gev,\;\;m_{\neutone}=439\gev,\qquad\qquad \textrm{light neutralino:}\;& m_{\tilde{d}_L}^{0}=1064\gev,\non\\
m_{\tilde{d}_R}^{\textrm{min}}=439\gev,\;\;m_{\neutone}=418\gev,\qquad\qquad \textrm{light neutralino:}\;& m_{\tilde{d}_R}^{0}=726\gev,\non\\
m_{\tilde{u}_L}^{\textrm{min}}=477\gev,\;\;m_{\neutone}=440\gev,\qquad\qquad \textrm{light neutralino:}\;&m_{\tilde{u}_L}^{0}=1061\gev,\non\\
m_{\tilde{u}_R}^{\textrm{min}}=456\gev,\;\;m_{\neutone}=323\gev,\qquad\qquad \textrm{light neutralino:}\;&m_{\tilde{u}_R}^{0}=902\gev,\non\\
\mgluino^{\textrm{min}}=792\gev,\;\;m_{\neutone}=661\gev,\qquad\qquad \textrm{light neutralino:}\;&\mgluino^{0}=1452\gev.
\eea

When comparing to the 8 TeV results one can notice that the minimal allowed gluino mass has increased by approximately 300\gev\ in the part of the parameter space where 
neutralino LSP is light, and by 250\gev\ in the more compressed region. 
In the case of the squarks, the corresponding increase of the minimal allowed mass is lower, in the range of $50-200$\gev. 
One should bare in mind, however, that since our recast procedure is an accurate, yet still approximate, simulation of the experimental analysis, the precise values of the minimal masses given above should be treated with some caution.

\section{Summary}\label{sum}

In this paper we perfomed the first analysis of the p19MSSM in light of 13 TeV LHC data with integrated luminosity 
of $~\sim14$\invfb. We recast seven ATLAS SUSY searches based on the 3.2\invfb\ data set and updated four of them, with the largest expected sensitivity, to incorporate new higher luminosity results.

We found that 25\% of the p19MSSM parameter space that was phenomenologically allowed after  inclusion of the 8 TeV data is now excluded at 95\% \cl, predominantly 
through all-hadronic multijet searches. In the part of the parameter space where neutralino is lighter than 50\gev, the lower limit on the gluino mass can be set at 
1450\gev, on the stop and sbottom masses at around 750\gev\ and 650\gev, respectively, and on the left squark masses at 1060\gev. 
In the compressed spectra region, which is notoriously difficult to test at the LHC as the decay products are soft, the new 95\% \cl\ exclusion bounds on the 
sparticle masses have reached approximately 790\gev\ (gluinos), 400\gev\ (stops), 350\gev\ (sbottoms), and 470\gev\ (left squarks). 

Some of the experimental ATLAS analyses based on $\sim14$ \invfb\ of data reported moderate excesses in the number of signal events with respect to the SM  background yields, reaching local significance of around $2-3\sigma$. One can hope these are harbingers of a future discovery. On the other hand, if none of the excesses turns into a SUSY signal when more data is collected, in the following months yet another chunk of the p19MSSM parameter space should be excluded by the LHC.

\bigskip  
\begin{center}
\textbf{ACKNOWLEDGMENTS}
\end{center}
I would like to thank Enrico Maria Sessolo for helpful discussions regarding the statistical treatment of the LHC limits.
This project is supported in part by the DFG Research Unit FOR 1873 ``Quark Flavour Physics and Effective Field
Theories''. 
The use of the CIS computer cluster at the National Centre for Nuclear Research in Warsaw is gratefully acknowledged. 
\bigskip
\newpage

\appendix

\section{Implemented 13 TeV searches}\label{13searches}
In this Appendix we present a summary of the implemented searches. We focus mainly on the consistency check  
between our recast and the official results by ATLAS. The technical details of each analysis can be found the in the experimental papers.
Note that in the validation analyses we used the NLO+NNL cross-sections, provided by LHC SUSY Cross Section Working Group\cite{LHCSXSECWG}.
\subsection{Search for squarks and gluinos in final states with jets and missing transverse momentum\cite{Aaboud:2016zdn,ATLAS-CONF-2016-078}.}
In this search a direct pair production of the gluinos and squark is analyzed. The experimental signature is composed of 2-6 hadronic jets and large amount of missing energy. The analysis performs the following pre-selection cuts:
\begin{itemize}
\item lepton veto with $p_T > 10\gev$,
\item at least two jets with $p_T > 50\gev$,
\item $\met > 200\gev$ (3.2 \invfb), 250\gev (13.3 \invfb).
\end{itemize}
Inclusive signal bins are defined for varying jet multiplicities and varying ranges of the following kinematical variables:
the leading jets transverse momentum $p_T(\textrm{jet})$, the minimal azimuthal separation $\mindfi$, the ratio of missing energy and a scalar sum of jets transverse momenta $\metht$, and $\meff$ defined as the scalar sum of the transverse momenta of the 
jets with $p_T>50$ GeV and $\met$. In \reftable{tab:ATLAS_0l26j_1} we show a cut flows comparison  between the experimental analysis 
by ATLAS with 3.2\invfb\ of data and the results of our recast, for a signal benchmark point $(\mgluino,m_{\neutone})=(1100,700)$ GeV. The same comparison for a point $(m_{\squark},m_{\neutone})=(1000,400)$ GeV is shown in \reftable{tab:ATLAS_0l26j_2}. The signal regions are presented that provide the most stringent exclusion.

\begin{table}[h]\footnotesize
\begin{centering}
\begin{tabular}{|c|c|c|c|c|c|c|}
\hline 
Cuts& $\met>200$ GeV, $\ptone$ & $N_{jet}$ & $\mindfi$ & $\pttwo$ & $\metht$  & $\meff$ \\
\hline 
SR2 - ATLAS & 35.1\% & 35.0\% & 29.3\% & 29.3\% & 11.8\% & 3.3\% \\
SR2 - recast & 36.5\% & 36.5\% & 29.8\% & 29.7\% & 11.6\% & 4.8\% \\
\hline
SR5 - ATLAS & 57.7\% & 26.8\% & 20.0\% & 19.7\% & 5.1\% & 1.5\% \\
SR5 - recast & 62.1\% & 33.1\% & 23.4\% & 23.1\% & 6.5\% & 2.7\% \\
\hline
\end{tabular}
\caption{\footnotesize Comparison of the cut flows for the signal point $(\mgluino,m_{\neutone})=(1100,700)$ GeV in the 0-lepton + 2-6 jets + $\met$ 
ATLAS 3.2\invfb\ search and in the recast tool.}
\label{tab:ATLAS_0l26j_1}
\end{centering}
\end{table}

\begin{table}[h]\footnotesize
\begin{centering}
\begin{tabular}{|c|c|c|c|c|c|c|}
\hline 
Cuts& $\met>200$ GeV, $\ptone$ & $N_{jet}$ & $\mindfi$ & $\pttwo$ & $\metht$  & $\meff$ \\
\hline 
SR2 - ATLAS & 77.0\% & 75.9\% & 67.8\% & 67.8\% & 44.5\% & 20.8\% \\
SR2 - recast & 77.6\% & 77.6\% & 68.6\% & 67.8\% & 45.3\% & 22.5\% \\
\hline
SR3 - ATLAS & 84.8\% & 83.5\% & 66.1\% & 49.3\% & 19.6\% & 3.8\% \\
SR3 - recast & 85.2\% & 85.2\% & 66.1\% & 47.3\% & 19.1\% & 5.2\% \\
\hline
\end{tabular}
\caption{\footnotesize Comparison of the cut flows for the signal point $(m_{\squark},m_{\neutone})=(1000,400)$ GeV in the 0-lepton + 2-6 jets + $\met$ ATLAS  3.2\invfb\ search and in the recast tool.}
\label{tab:ATLAS_0l26j_2}
\end{centering}
\end{table}

In \reffig{fig:ATLAS_0l26j}\subref{fig:a} we present a validation of our simulation in terms of the exclusion limits in the parameter space of $(\mgluino,m_{\neutone})$ for 3.2 \invfb\ analysis.
Gray diamonds represent the points excluded by our likelihood function at the 99.7\%~C.L., 
cyan circles are excluded at the 95.0\%~C.L., 
and blue triangles are excluded at the 68.3\%~C.L. 
The points depicted as red squares are considered as allowed. 
The solid black line shows the 95\%~C.L. ATLAS exclusion limit, which we present for comparison.  
A corresponding validation in the $(m_{\squark},m_{\neutone})$ plane is shown in \reffig{fig:ATLAS_0l26j}\subref{fig:b}.
Validations of the 13.3 \invfb\ analysis, for the same SMS, are shown in \reffig{fig:ATLAS_0l26j}\subref{fig:c} and \reffig{fig:ATLAS_0l26j}\subref{fig:d}.

\begin{figure}[t]
\centering
\subfloat[]{
\label{fig:a}
\includegraphics[width=0.40\textwidth]{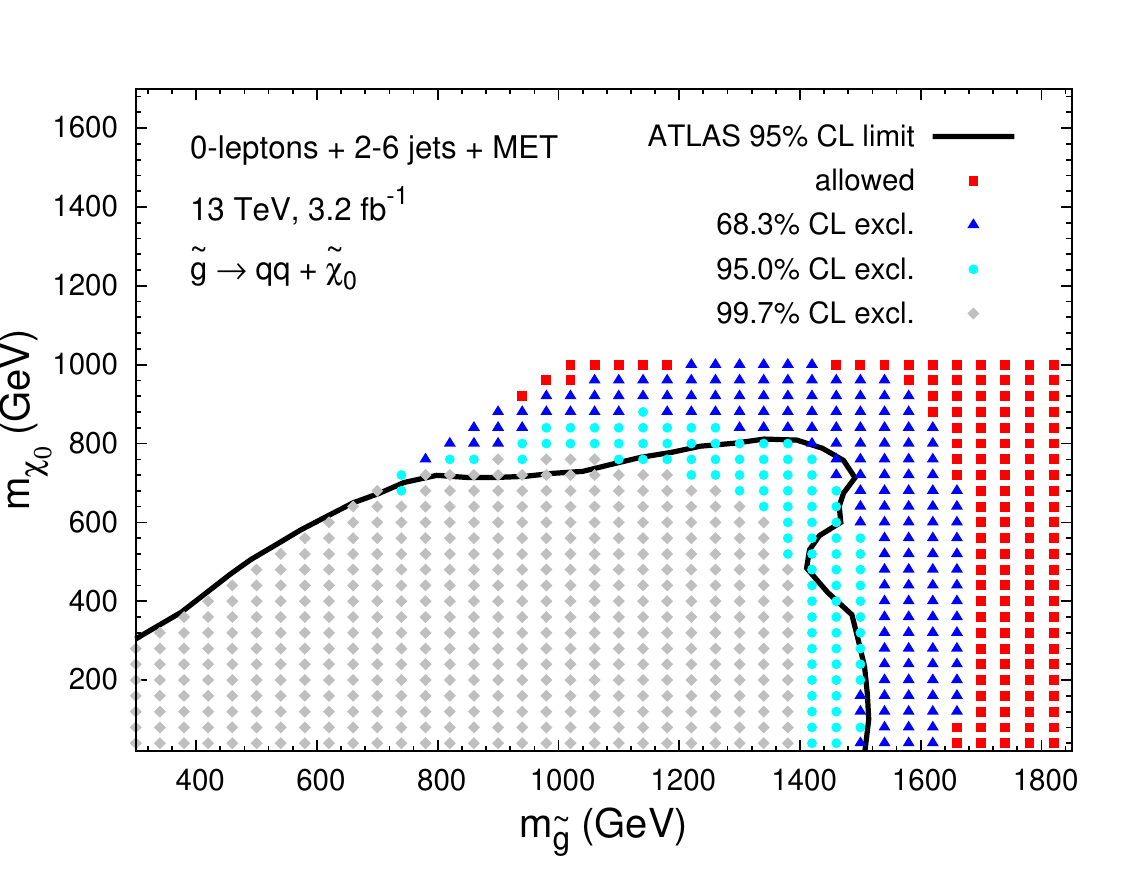}
}
\subfloat[]{
\label{fig:b}
\includegraphics[width=0.40\textwidth]{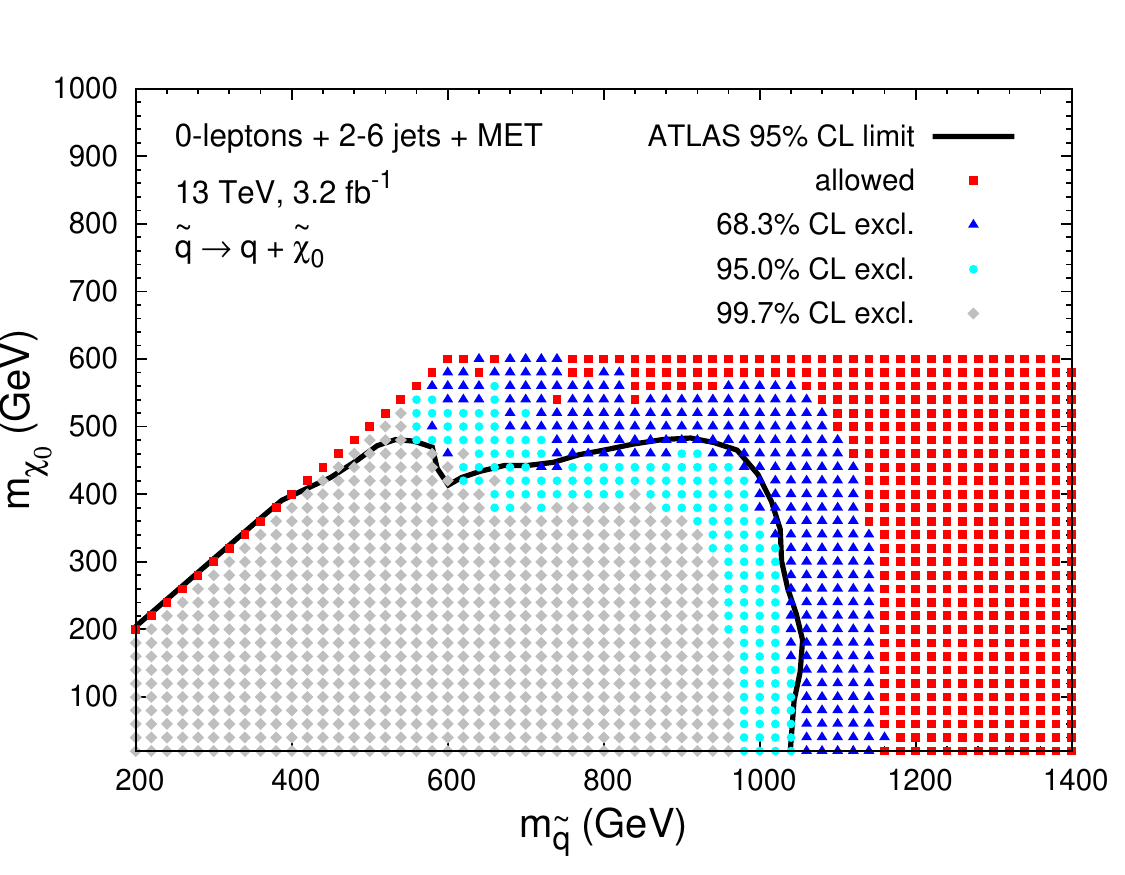}
}\\
\subfloat[]{
\label{fig:c}
\includegraphics[width=0.40\textwidth]{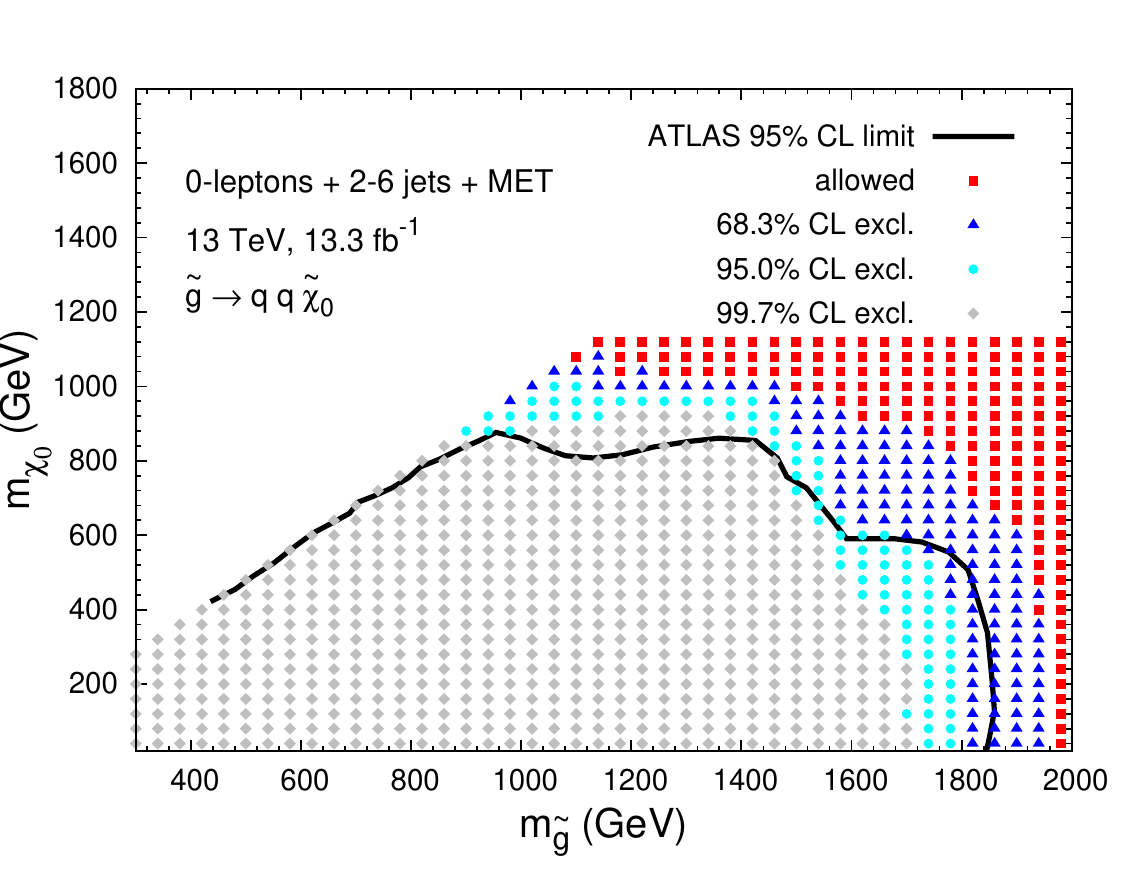}
}
\subfloat[]{
\label{fig:d}
\includegraphics[width=0.40\textwidth]{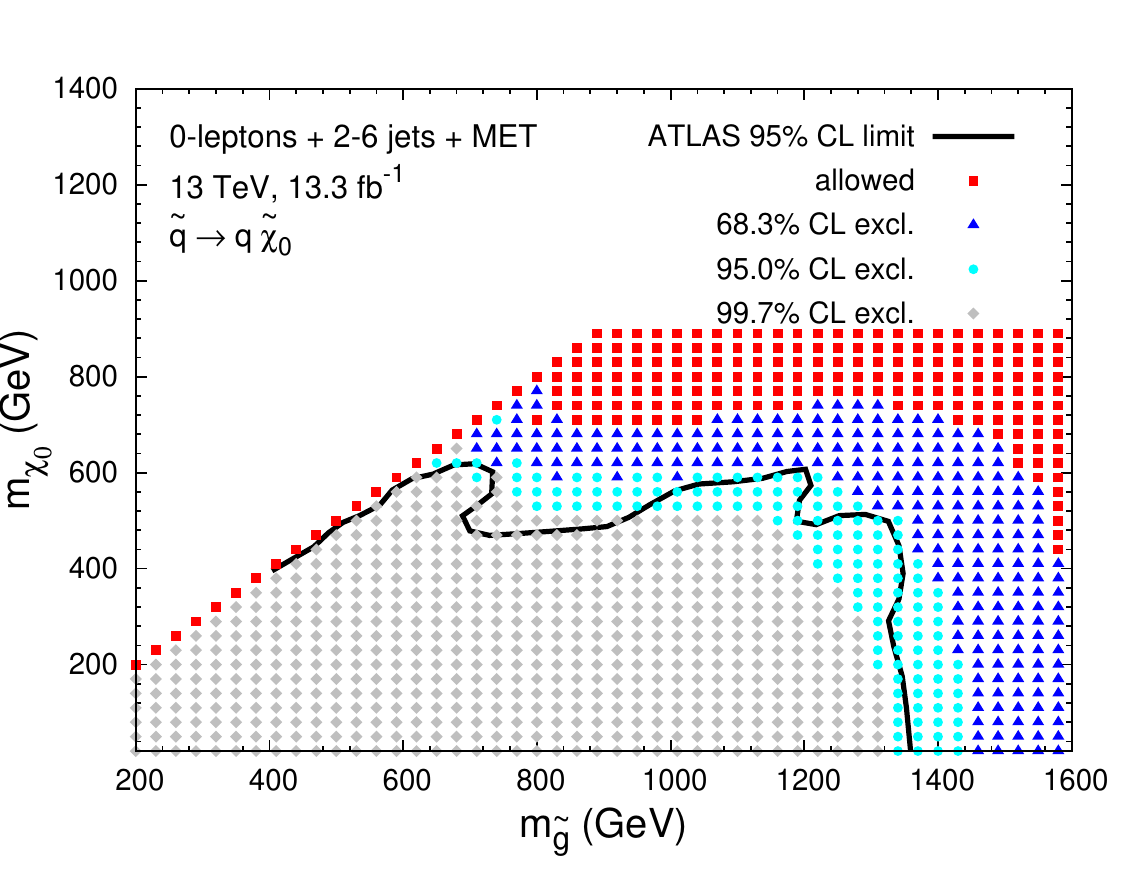}
}
\caption[]{\footnotesize \subref{fig:a} Our simulation of the ATLAS 0-lepton search with 3.2 \invfb\ of data for direct gluino production assuming a decay chain $\tilde{g}\to qq\neutone$. 
\subref{fig:b} The same for direct squark production assuming a decay chain $\tilde{q}\to q\neutone$.  
\subref{fig:c} and  \subref{fig:d} The same for 13.3 \invfb\ analysis. 
Points that are excluded at the 99.7\%~C.L. are shown as gray diamonds, at the 95.0\%~C.L. as cyan circles, 
and at the 68.3\%~C.L. as blue triangles. The points shown as red squares are considered as allowed. 
The solid black lines show the published 95\%~C.L. contours by ATLAS.}
\label{fig:ATLAS_0l26j}
\end{figure}

\subsection{Search for gluinos in events with an isolated lepton, jets and missing transverse momentum\cite{Aad:2016qqk,ATLAS-CONF-2016-054}.}

In this search a direct pair production of gluinos is assumed that undergo a 3-body decay into two quarks and chargino. The latter subsequently
decays into neutralino LSP and the on/off-shell $W$ boson. The experimental signature is characterized by presence of exactly one lepton, hadronic jets
and large amount of missing energy. The analysis performs the following pre-selection cuts:
\begin{itemize}
\item one electron(muon) with $p_T > 7(6)\gev$,
\item at least two jets with $p_T > 30\gev$.
\end{itemize}
\begin{table}[h]\footnotesize
\begin{centering}
\begin{tabular}{|c|c|c|c|c|c|c|}
\hline 
Cuts& $\met$ & $p_T(jet_1)$ & $p_T(jet_\textrm{last})$ & $m_T$ & $\metmef$  & $\meff$ \\
\hline 
5-jet SR - ATLAS & 19.2\% & 18.2\% & 16.0\% & 8.5\% & 8.3\% & 7.6\% \\
5-jet SR - recast & 18.2\% & 17.4\% & 15.3\% & 9.6\% & 9.5\% & 8.3\% \\
\hline
6-jet SR - ATLAS & 19.2\% & 15.1\% & 15.1\% & 9.0\% & 5.3\% & 5.3\% \\
6-jet SR - recast & 18.2\% & 16.3\% & 15.3\% & 10.7\% & 6.6\% & 6.6\% \\
\hline
4-jet high-x SR - ATLAS & 21.6\% & 19.4\% & 19.4\% & 19.4\% & 1.5\% & 1.4\% \\
4-jet high-x SR - recast & 19.7\% & 16.5\% & 16.5\% & 16.5\% & 2.8\% & 2.3\% \\
\hline
\end{tabular}
\caption{\footnotesize Comparison of the cut flows for the signal point $(\mgluino,m_{\charone},m_{\neutone})=(1385,705,25)$ GeV in 
the 1-lepton + jets + $\met$ ATLAS 3.2\invfb\ search and in the recast tool.}
\label{tab:ATLAS_1ljet}
\end{centering}
\end{table}
The inclusive signal bins are defined for varying jet multiplicities, soft- and hard-lepton channels (with $p_T^l<35$ \gev\ and $p_T^l\geq35$, respectively)
and varying ranges of the following kinematical variables:
the leading jets transverse momentum $p_T(\textrm{jet})$,  the transverse mass $m_T$ of the signal lepton and missing transverse momentum, 
sum of transverse momenta of all signal jets and the signal lepton $H_T$, and the effective mass $\meff$ defined as a sum over 
transverse momenta of $\met$, all signal jets and the lepton.
In \reftable{tab:ATLAS_1ljet} we show a  cut flows comparison between the experimental analysis by ATLAS with 3.2 \invfb\ of data and the results of our recast,
for a signal benchmark point $(\mgluino,m_{\charone},m_{\neutone})=(1385,705,25)$ GeV. 
In \reffig{fig:ATLAS_1ljet} we present a validation of our simulation in terms of the exclusion limits in the parameter space of 
$(\mgluino,m_{\neutone})$. \reffig{fig:ATLAS_1ljet}\subref{fig:a} corresponds to the 3.2 \invfb\ analysis, while \reffig{fig:ATLAS_1ljet}\subref{fig:b} to its 14.8 \invfb\ update. The colour code is the same as in \reffig{fig:ATLAS_0l26j}.

\begin{figure}[b]
\centering
\subfloat[]{
\label{fig:a}
\includegraphics[width=0.40\textwidth]{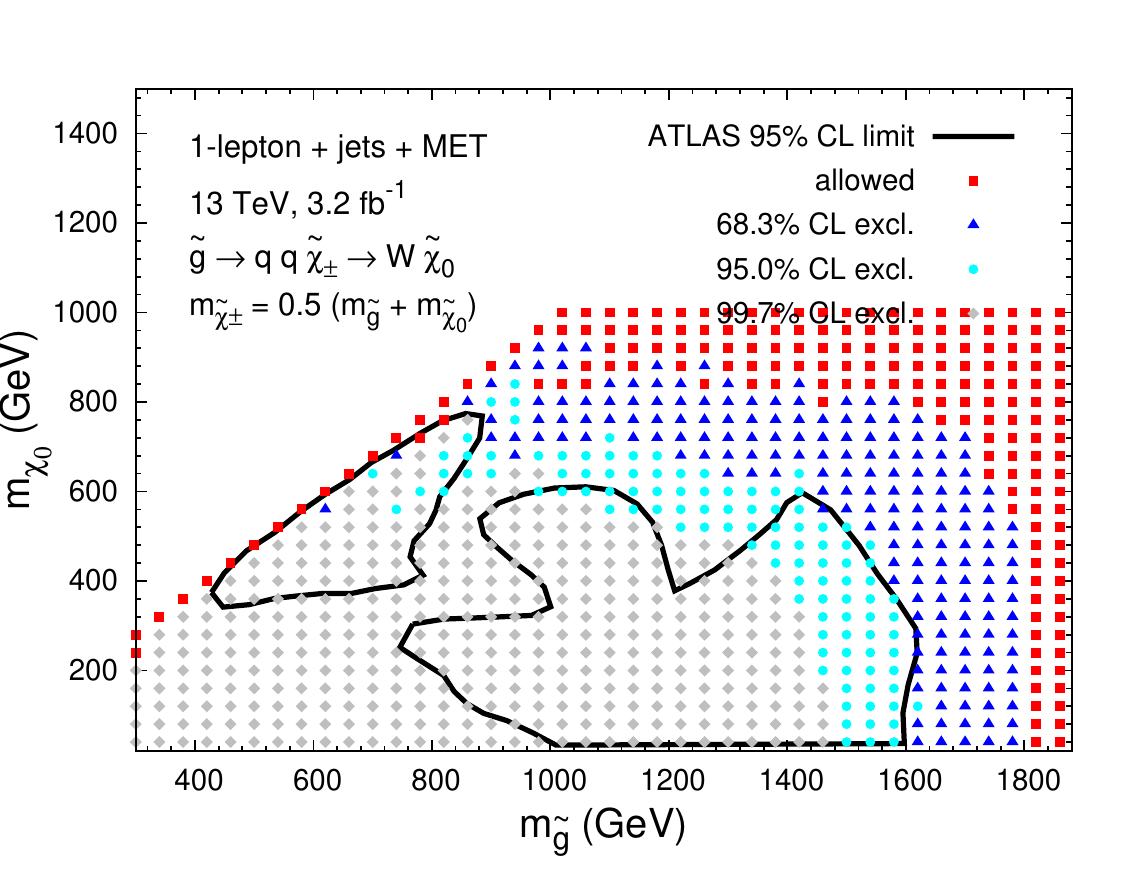}
}
\subfloat[]{
\label{fig:b}
\includegraphics[width=0.40\textwidth]{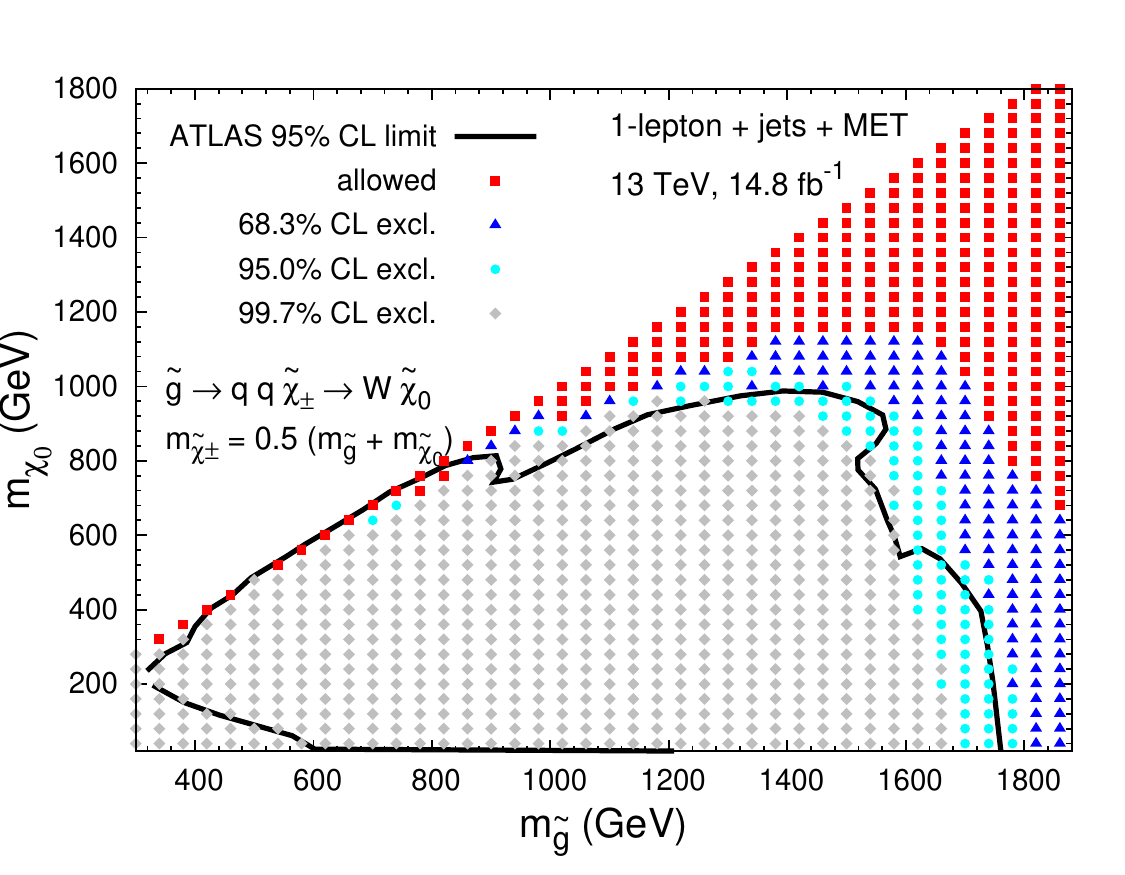}
}
\caption[]{\footnotesize \subref{fig:a} Our simulation of the ATLAS 1-lepton search in the 3.2 \invfb\ analysis for direct gluino production for the mass of chargino $m_{\charone}=\frac{1}{2}(\mgluino+m_{\neutone})$. 
\subref{fig:b} The same for 14.8 \invfb\ analysis.
The colour code is the same as in \reffig{fig:ATLAS_0l26j}.}
\label{fig:ATLAS_1ljet}
\end{figure}
\subsection{Search for pair production of gluinos decaying via stop and sbottom in events with b-jets and large missing transverse 
momentum\cite{Aad:2016eki,ATLAS-CONF-2016-052}.}

In this search a direct pair production of the gluinos is assumed which can than decay through the off-shell sbottoms and stops. 
The experimental signature is characterized by at least three energetic b-tagged jets and large amount of missing energy. The analysis performs the following pre-selection cuts:
\begin{itemize}
\item at least 4 signal jets with $p_T>30$ GeV,
\item at least 3 b-jets with $p_T>30$ GeV,
\item $\met > 200\gev$.
\end{itemize}
The following kinematical variables are used to discriminate between the signal and the background: the effective mass $\meff$ defined as a sum over 
transverse momenta of $\met$, all signal jets and leptons; its subclass $\meffj$ which includes four leading jests only; the minimum transverse mass 
$\mtb$ formed by missing energy and any of the three leading b-tagged jets; and the minimum
azimuthal angle  $\dfij$ between $\met$ and the leading four jets.

\begin{table}[t]\footnotesize
\begin{centering}
\begin{tabular}{|c|c|c|c|c|c|c|}
\hline 
Cuts& $\geq 4$ jets & $\geq 3$ b-jets &  $\dfij$ & $\meffj$/$\meff$  & $\met>350$ GeV & $p_T(\textrm{jet})>90$ GeV\\
\hline 
Gbb-A - ATLAS & 94.8\% & 59.0\% & 39.6\% & 36.6\% & 33.2\%  & 25.5\%\\
Gbb-A - recast & 98.6\% & 52.3\% & 35.4\% & 32.2\% & 29.0\% & 22.2\%\\
\hline
Gtt-0L-A - ATLAS & 96.3\% & 71.8\% & 26.4\% & 18.6\% & 15.4\% & -\\
Gtt-0L-A - recast & 99.9\% & 68.3\% & 28.0\% & 20.1\% & 16.6\% & - \\
\hline
\end{tabular}
\caption{\footnotesize Comparison of the cut flows for the signal point $(\mgluino,m_{\neutone})=(1700,200)$ GeV in the 3 b-jets + $\met$ 
ATLAS 3.2\invfb\ search and in the recast tool.}
\label{tab:ATLAS_3bj}
\end{centering}
\end{table}

\begin{figure}[b]
\centering
\subfloat[]{
\label{fig:a}
\includegraphics[width=0.40\textwidth]{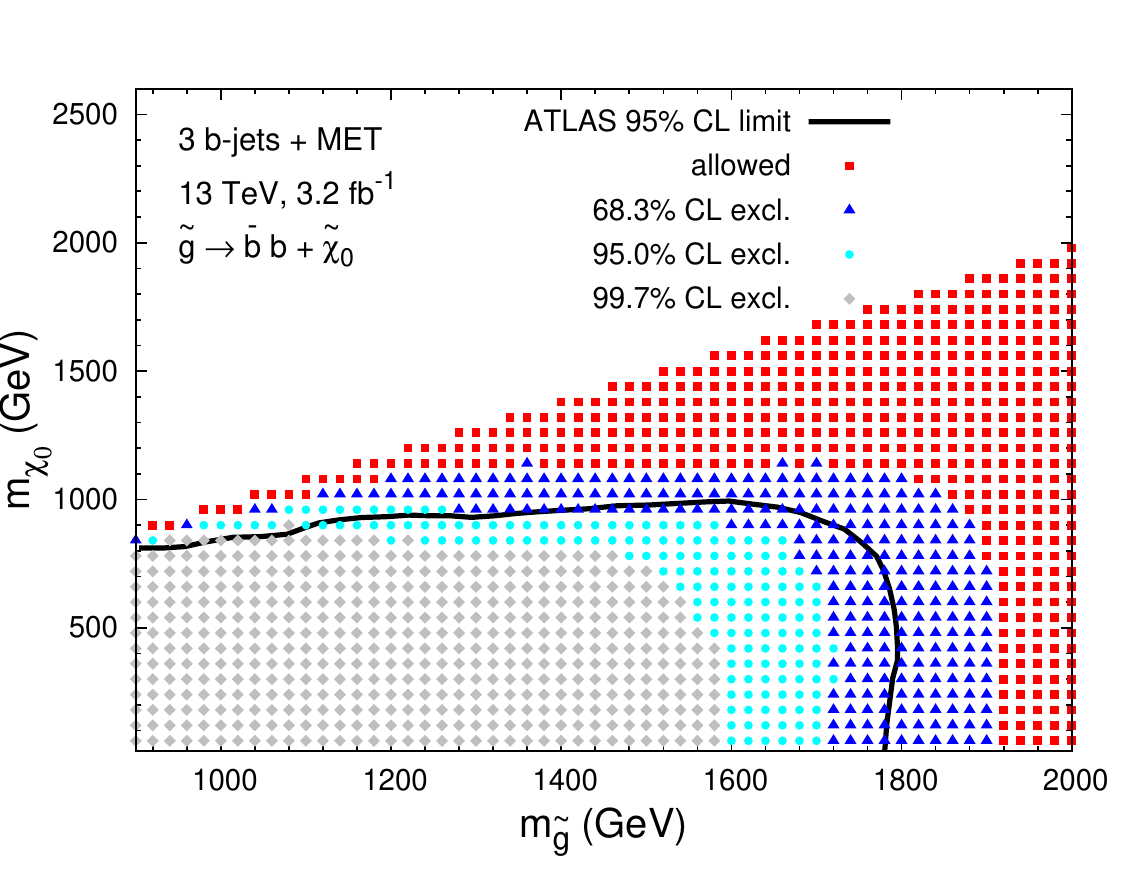}
}
\subfloat[]{
\label{fig:b}
\includegraphics[width=0.40\textwidth]{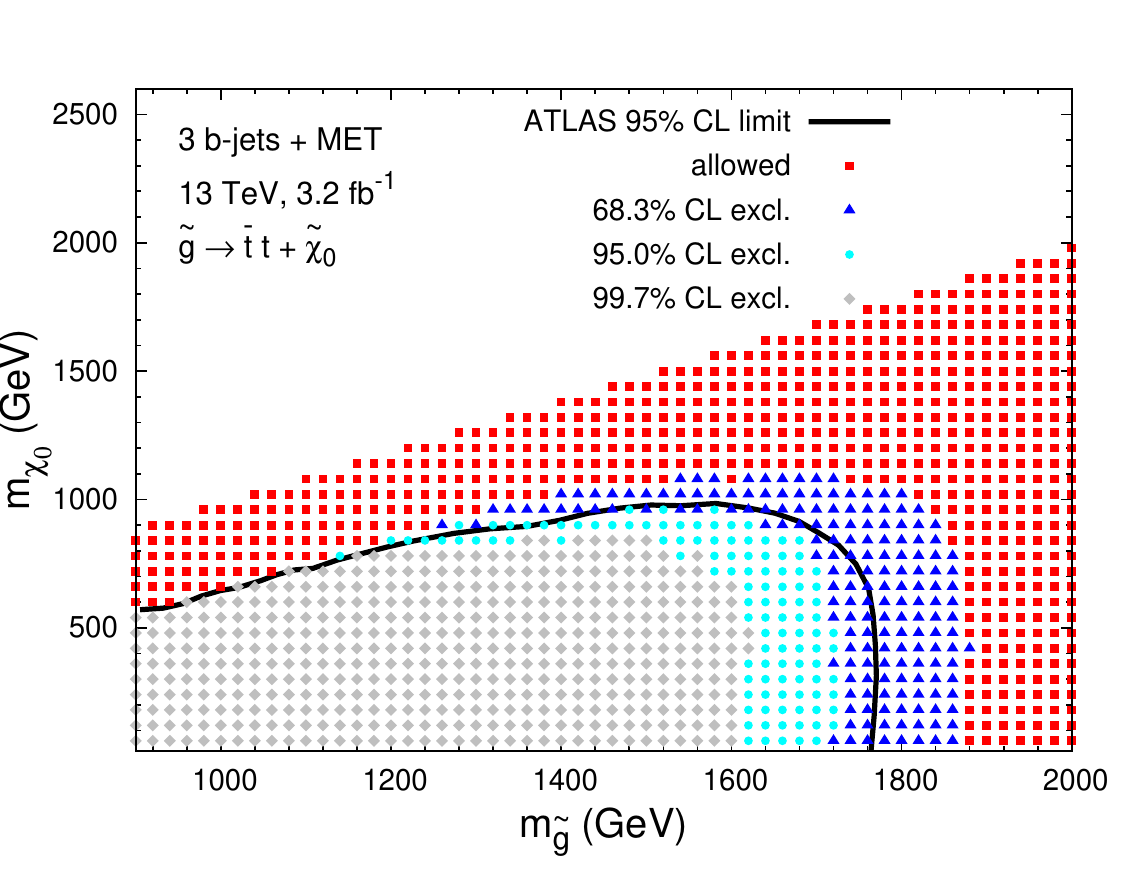}
}\\
\subfloat[]{
\label{fig:c}
\includegraphics[width=0.40\textwidth]{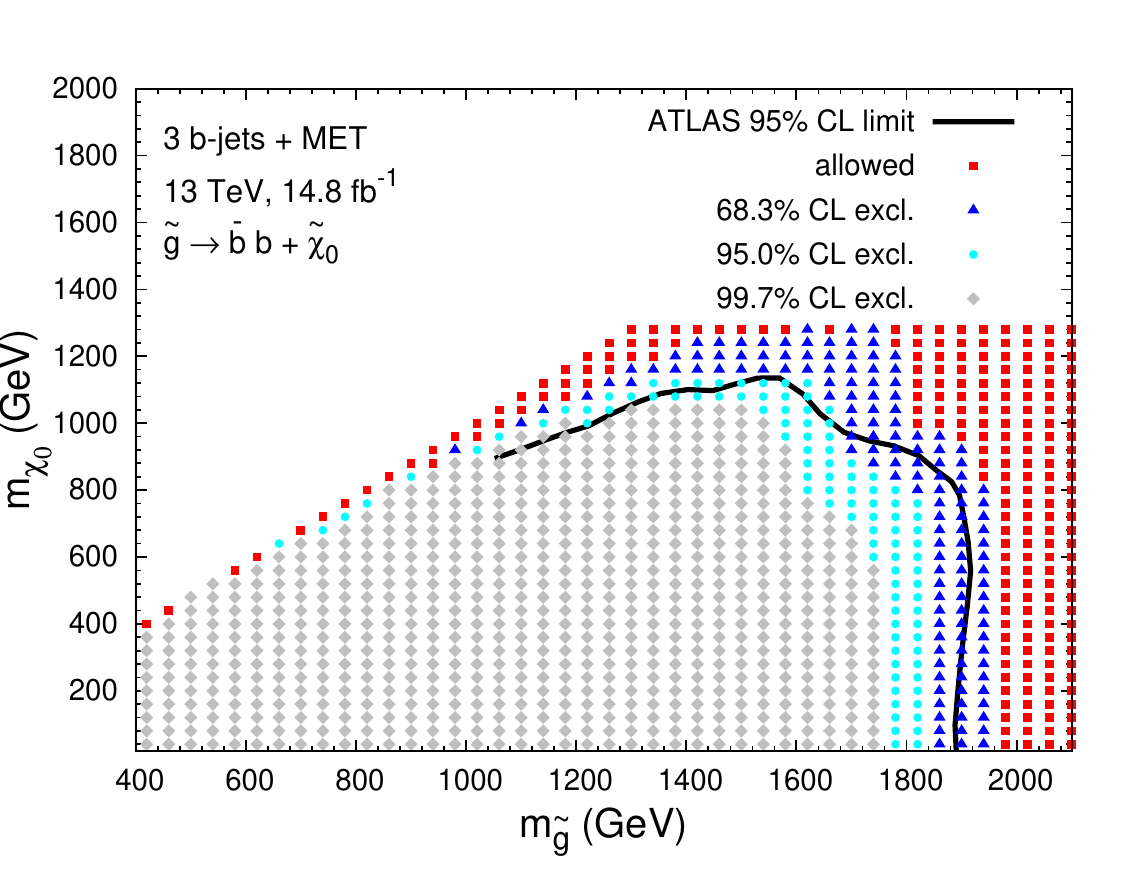}
}
\subfloat[]{
\label{fig:d}
\includegraphics[width=0.40\textwidth]{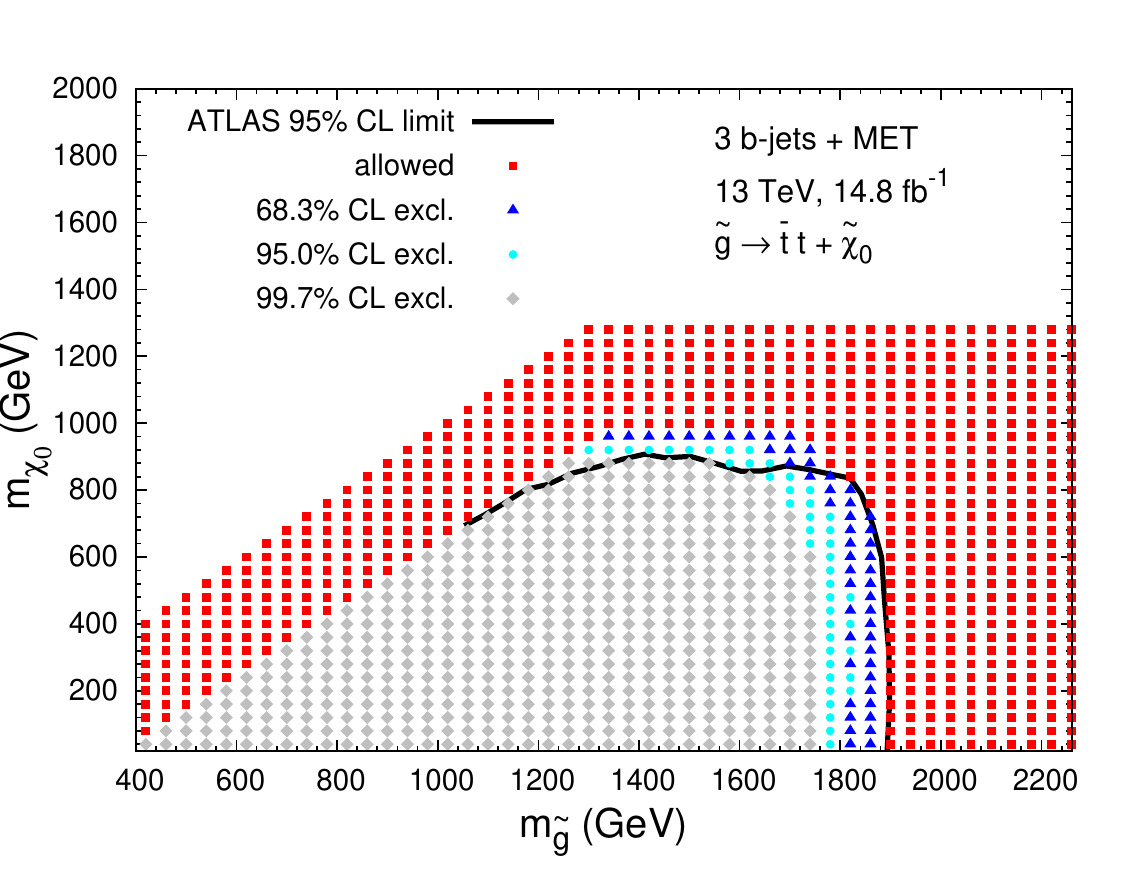}
}
\caption[]{\footnotesize \subref{fig:a} Our simulation of the ATLAS 3 b-tagged jets search with 3.2 \invfb\ of data for direct gluino production assuming a decay chain $\tilde{g}\to bb\neutone$. 
\subref{fig:b} The same for a decay chain $\tilde{g}\to tt\neutone$.  
\subref{fig:c} and \subref{fig:d} The same for the 14.8 \invfb\ analysis.
The colour code is the same as in \reffig{fig:ATLAS_0l26j}.}
\label{fig:ATLAS_3bj}
\end{figure}

For the sbottom-mediated decay model (denoted as Gbb) the inclusive signal regions are defined for varying ranges of missing energy and $\meffj$.
For the top-mediated decay model (denoted as Gtt)  inclusive signal regions with no leptons and with one lepton are defined, differentiated by 
the values of $\met$, $\meff$ and the number of b-tagged jets.

In \reftable{tab:ATLAS_3bj} we show a comparison of the cut flows between the experimental analysis by ATLAS with 3.2\invfb\ of data and the results of our recast, 
for a signal benchmark point $(\mgluino,m_{\neutone})=(1700,200)$ GeV. The signal regions with the best efficiency are presented.
In \reffig{fig:ATLAS_3bj}\subref{fig:a} we show a validation of our simulation in terms of the exclusion limits in the parameter space of 
$(\mgluino,m_{\neutone})$ for Gbb model. A corresponding validation for model Gtt is shown in \reffig{fig:ATLAS_3bj}\subref{fig:b}. 
Validations for the 14.8 \invfb\ analysis, for the same SMS, are shown in \reffig{fig:ATLAS_3bj}\subref{fig:c} and \reffig{fig:ATLAS_3bj}\subref{fig:d}.
The colour code is the same as in \reffig{fig:ATLAS_0l26j}.
\subsection{Search for supersymmetry in events containing a leptonically decaying Z boson, jets and missing transverse momentum\cite{ATLAS-CONF-2015-082}.}

In this search a pair production of gluinos is assumed. A simplified model considered here assumes a tree-body decay of gluino into quarks 
and second neutralino, with a subsequent decay of the latter into a neutralino and Z boson (reconstructed from two leptons). 
\begin{itemize}
\item at least two signal jest with $p_T>30$ GeV,
\item at least two signal leptons with $p_T>25$ GeV,
\item the leading and sub-leading leptons form a same-flavour opposite-sign pair,
\item $\met > 225\gev$.
\end{itemize}
The following kinematical variables are used to discriminate between the signal and the background: 
the invariant mass $m_{ll}$ of the dilepton system, and the sum of transverse momenta of all signal jets and two leading leptons $H_T$.
Only one signal region is defined with $H_T>600\gev$ and $81\gev <m_{ll}<101\gev$.

In \reffig{fig:ATLAS_1lj}\subref{fig:a} we present a validation of our simulation in terms of the exclusion limits in the parameter space of $(\mgluino,m_{\neuttwo})$. The colour code is the same as in \reffig{fig:ATLAS_0l26j}. For the second neutralino mass up to 400\gev\ we obtain a very good agreement with the experimental bound. For larger values of $m_{\neuttwo}$ a discrepancy is observed that reaches around 100 \gev\ for $m_{\neuttwo}\sim 800\gev$.

\begin{figure}[b]
\centering
\subfloat[]{
\label{fig:a}
\includegraphics[width=0.40\textwidth]{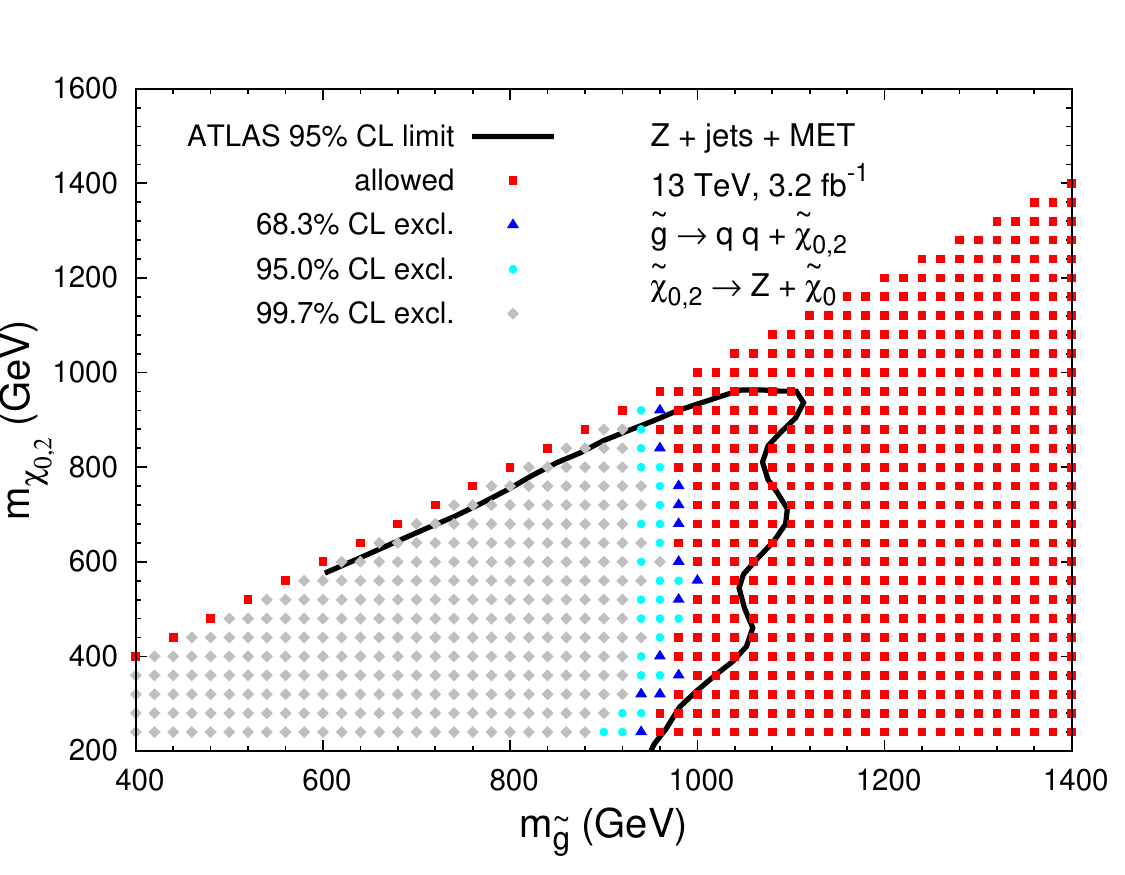}
}
\subfloat[]{
\label{fig:b}
\includegraphics[width=0.40\textwidth]{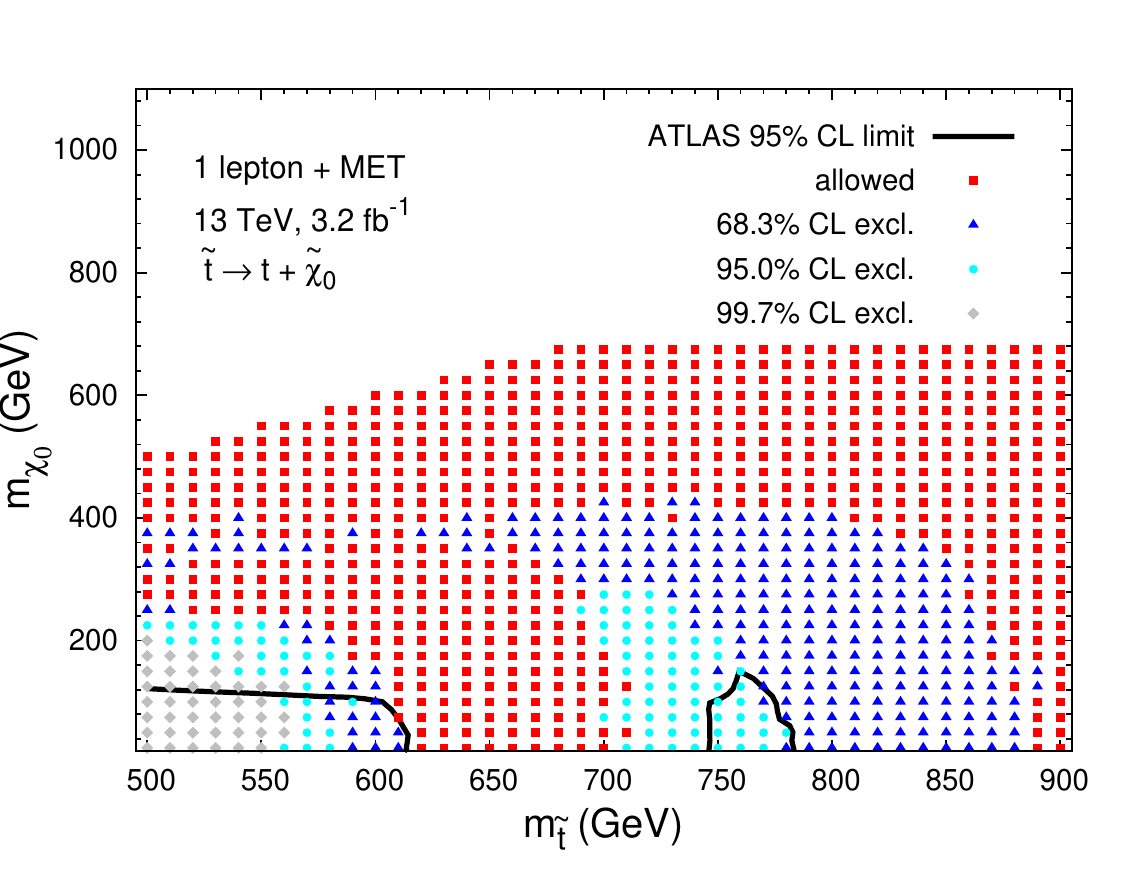}
}
\caption[]{\footnotesize \subref{fig:a} Our simulation of the ATLAS Z + jets search for direct gluino production. 
\subref{fig:b} The same for ATLAS 1-lepton search for direct stop production and a decay $\tilde{t}_1\to t \neutone$. Both validations correspond to 3.2 \invfb\ analyses.
The colour code is the same as in \reffig{fig:ATLAS_0l26j}.}
\label{fig:ATLAS_1lj}
\end{figure}
\subsection{Search for top squarks in final states with one isolated lepton, jets, and missing transverse 
momentum\cite{Aaboud:2016lwz,ATLAS-CONF-2016-050}.}

In this search a light partner of the top quarks is tested in two scenarios: gluino-mediated pair
production of the stop with a small stop and neutralino LSP mass splitting, and direct pair production of the stops.
The experimental signature includes one lepton, jets and large amount of missing energy. The analysis performs the following pre-selection cuts:
\begin{itemize}
\item exactly one signal lepton with $p_T>25$ GeV,
\item at least 4 signal jets with $p_T>25$ GeV,
\item $\met > 200\gev$.
\end{itemize}
The following kinematical variables are used to discriminate between the signal and the background: 
azimuthal angle $\dfi$ between two leading jets and missing energy, the transverse mass $m_T$ of the signal lepton and missing transverse momentum, the asymmetric stransverse mass $\amt$, the invariant mass $\mctop$ of the three jets in
the event most compatible with the hadronic decay products of a top quark, and the angular separation $\delR$ between the signal lepton and the highest-$p_T$ b-jet.

The signal regions are defined to cover various decay topologies and kinematical regimes.
In \reftable{tab:ATLAS_1lj} we show a comparison of the cut flows between the experimental analysis by ATLAS with 3.2 \invfb\ of data and the results of our recast, 
for a signal benchmark point $(\mtone,m_{\neutone})=(600,260)$ GeV in the SR1 which is the most sensitive bin in this scenario. 

\begin{table}[t]\footnotesize
\begin{centering}
\begin{tabular}{|c|c|c|c|c|c|c|c|c|c|}
\hline 
Cuts& 1 lepton & $\geq 4$ jets &  $\dfi$ & $p_T(\textrm{jet}_i)$  & $\met$& $m_T$ & $\amt$ & $\mctop$ & $\delR$\\
\hline 
SR1 - ATLAS & 23.4\% & 16.1\% & 10.1\% & 7.0\% & 5.1\%  & 2.9\% & 2.1\% & 1.7\% & 1.6\%\\
SR1 - recast & 20.2\% & 15.3\% & 8.9\% & 6.0\% & 4.3\% & 2.9\% & 2.6\% & 1.9\% & 1.8\%\\
\hline
\end{tabular}
\caption{\footnotesize Comparison of the cut flows for the signal point $(\mtone,m_{\neutone})=(600,260)$ GeV in the 1-lepton + jets + $\met$ 
ATLAS 3.2 \invfb\ search and in the recast tool.}
\label{tab:ATLAS_1lj}
\end{centering}
\end{table}

\begin{figure}[b]
\centering
\subfloat[]{
\label{fig:a}
\includegraphics[width=0.40\textwidth]{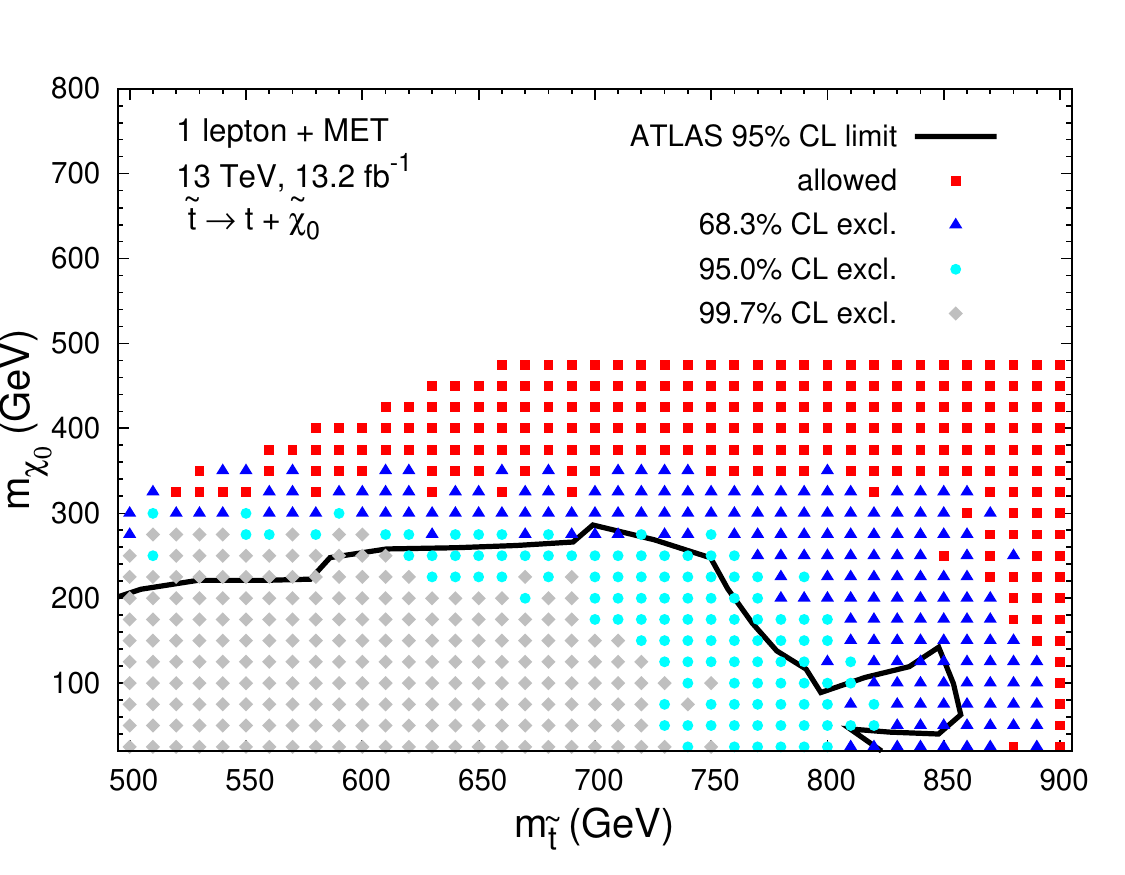}
}
\subfloat[]{
\label{fig:b}
\includegraphics[width=0.40\textwidth]{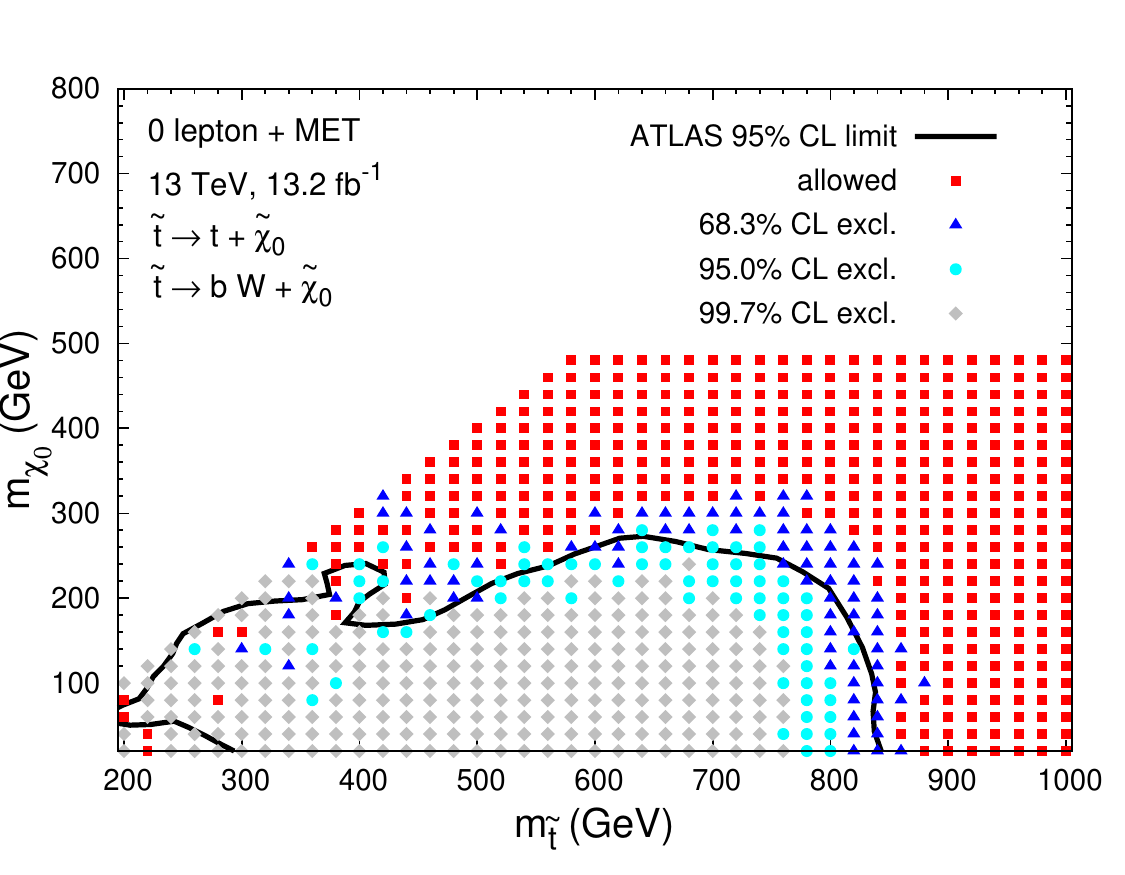}
}
\caption[]{\footnotesize \subref{fig:a} Our simulation of the ATLAS 1-lepton search with 13.2 \invfb\ of data for direct stop production and a decay $\tilde{t}_1\to t \neutone$. 
\subref{fig:b} The same for the corresponding ATLAS 0-lepton analysis.
The colour code is the same as in \reffig{fig:ATLAS_0l26j}.}
\label{fig:ATLAS_1ljnew}
\end{figure}


In \reffig{fig:ATLAS_1lj}\subref{fig:b} we present a validation of our simulation in terms of the exclusion limits in the parameter space of $(\mtone,m_{\neutone})$ for the direct stop production scenario in the analysis based on 3.2 \invfb\ of data. A corresponding validation for the 13.2 \invfb\ search is shown in \reffig{fig:ATLAS_1ljnew}\subref{fig:a}. The colour code is the same as in \reffig{fig:ATLAS_0l26j}.
\subsection{Search for the supersymmetric partner of the top quark in the jets plus missing energy final state\cite{ATLAS-CONF-2016-077}.}

In this search a pair-production of the light partner of the top quarks is tested. 
The experimental signature includes jets (two of which are tagged as originating from a b-quark) and large amount of missing energy. The analysis performs the following pre-selection cuts:
\begin{itemize}
\item lepton veto,
\item at least 4 signal jets with $p_T>40$ GeV,
\item $\met > 250\gev$.
\end{itemize}
The following kinematical variables are used to discriminate between the signal and the background: 
azimuthal angle $\dfi$ between two leading jets and missing energy, the transverse mass $m_T^b$ of the  missing transverse momentum and the b-tagged jet closest in $\phi$ to $p_T^{\textrm{miss}}$, the angular separation $\Delta R(b,b)$ between two b-tagged jets, the ratio of missing energy and a scalar sum of jets transverse momenta $\metht$,  and the mass $m^{0,1}_{\textrm{jet},R=1.2}$ of the two leading jets reclustered with a distance parameter $R=1.2$

Eleven signal regions are defined to target various kinemtaic regimes.
In \reffig{fig:ATLAS_1ljnew}\subref{fig:b} we present a validation of our simulation in terms of the exclusion limits in the parameter space of $(\mtone,m_{\neutone})$ for the direct stop production scenario in the analysis based on the 13.2 \invfb\ of data. The SMS is tested in which stop decays via $\tilde{t}_1\to t \neutone$ or  $\tilde{t}_1\to bW \neutone$, depending on kinematics. The colour code is the same as in \reffig{fig:ATLAS_0l26j}.

\subsection{Search for direct top squark pair production in final states with two leptons\cite{ATLAS-CONF-2016-009}.}
In this search a pair production of the top squarks is assumed, decaying softly into chargino close in mass, $\tone\to b\charone$, which then 
undergoes a decay $\charone\to W\neutone$. The experimental signature is characterized by exactly two opposite charge leptons and large amount of missing energy. 
The analysis performs the following pre-selection cuts:
\begin{itemize}
\item leading lepton with $p_T>25$ GeV and subleading lepton with $p_T>15$ GeV,
\item no other leptons with $p_T>10$ GeV.
\end{itemize}
The following kinematical variables are used to discriminate between the signal and the background: 
the invariant mass of the two leptons $m_{ll}$, the leptonic stransverse mass $\mtt$, and the ratio $\met/\meff$, where $\meff$ is
defined as the scalar sum of $\met$ and the momenta of leptons and up to two jets with $p_T>50$ GeV.

In the 3.2\invfb\ analysis, two exclusive signal regions are defined, one for the same flavour lepton pair, and the other for the opposite flavour
lepton pair. The exclusion limit is derived as a statistical combination of the limits from both regions.  In \reffig{fig:ATLAS_2bl}\subref{fig:a} we present a validation of our simulation in terms of the exclusion limits in the parameter space of $(\mtone,m_{\neutone})$, assuming $m_{\charone}=m_{\tone}-10$ GeV.

\begin{figure}[t]
\centering
\subfloat[]{
\label{fig:a}
\includegraphics[width=0.40\textwidth]{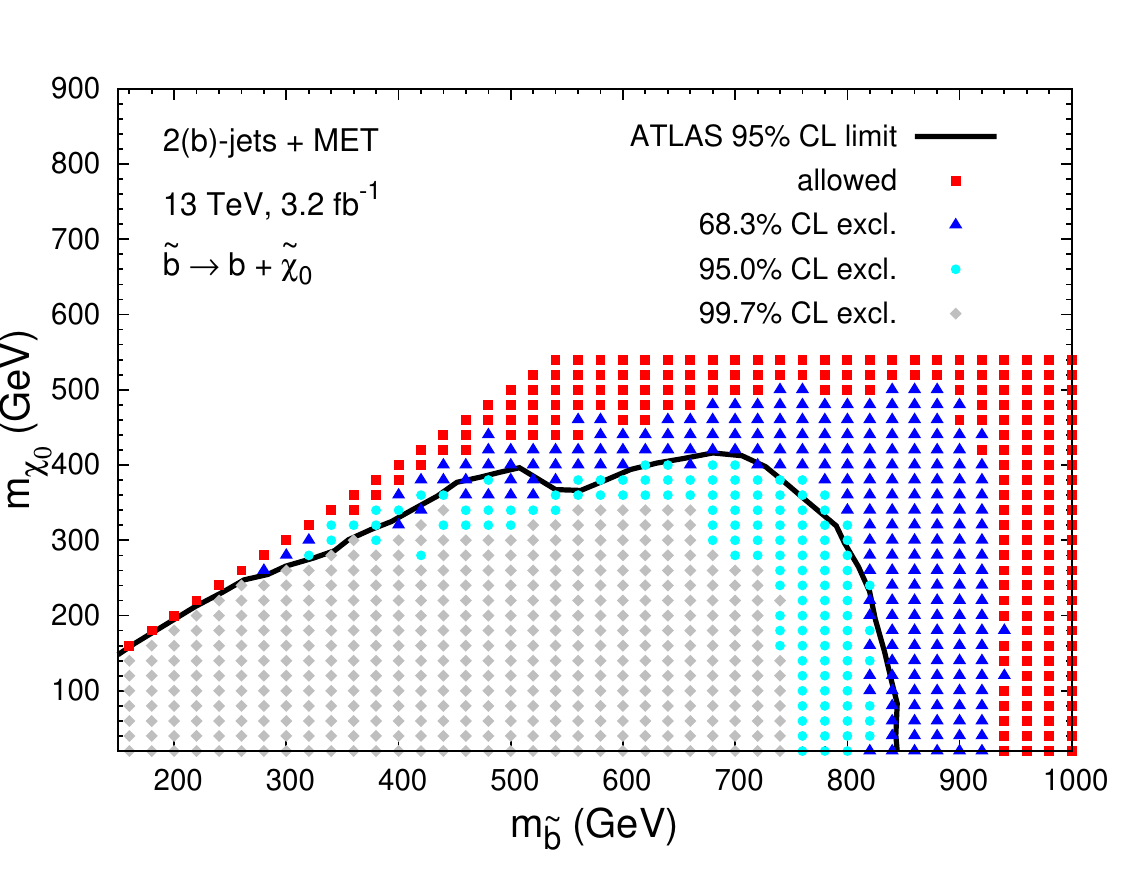}
}
\subfloat[]{
\label{fig:b}
\includegraphics[width=0.40\textwidth]{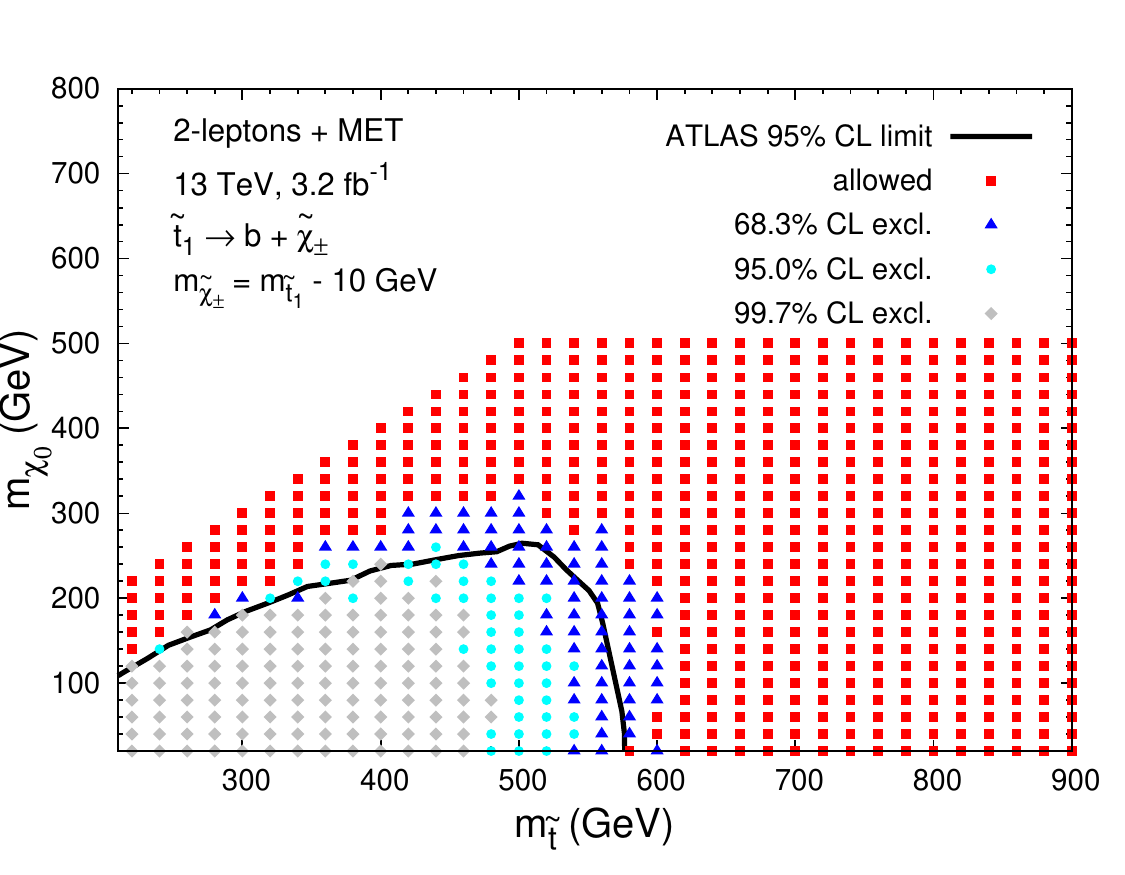}
}
\caption[]{\footnotesize \subref{fig:a} Our simulation of the ATLAS 2-lepton search for direct stop production.
\subref{fig:b} The same for ATLAS 2-b-jets search for direct sbottom production.  
The colour code is the same as in \reffig{fig:ATLAS_0l26j}.}
\label{fig:ATLAS_2bl}
\end{figure}
\subsection{Search for bottom squark pair production\cite{Aaboud:2016nwl}.}
In this search a pair production of the bottom squarks is assumed. 
The experimental signature includes two b-tagged jets and large amount of missing energy. The analysis performs the following pre-selection cuts:
\begin{itemize}
\item lepton veto with $p_T>10$ GeV,
\item $\met > 250\gev$.
\end{itemize}
The following kinematical variables are used to discriminate between the signal and the background: 
azimuthal angle $\dfi$ between the leading jet and missing energy, the minimum
azimuthal angle  $\dfij$ between $\met$ and the leading four jets, the invariant mass of the two b-jets $m_{bb}$, 
the contransverse mass $m_{\textrm{CT}}$, and the effective mass $\meff$ defined as the scalar sum of the $\met$ and the momenta the two (three) leading jets.

Two signal regions are defined, characterized either by two energetic b-jets or two b-jets accompanied by an energetic jet from the
initial state radiation. The exclusion limit is set based on the best SR at each point of the parameter space. 
In \reffig{fig:ATLAS_2bl}\subref{fig:b} we present a validation of our simulation in terms of the exclusion limits in the parameter space of 
$(\mbone,m_{\neutone})$.

\subsection{Search for new phenomena in final states with an energetic jet and large missing transverse momentum\cite{Aaboud:2016tnv}.}

\begin{table}[b]\footnotesize
\begin{centering}
\begin{tabular}{|c|c|c|c|c|c|}
\hline 
Cuts& lepton veto & $N_{\textrm{jets}}\leq4$ & $p_T(\textrm{jet}_1)>250$ \gev\ &  $\met>250$\gev  & $\met$\\
\hline 
EM1 - ATLAS & 79.9\% & 74.5\% & 17.8\% & 16.2\% & 3.2\%\\
EM1 - recast & 99.3\% & 69.3\% & 13.8\% & 11.7\% & 1.1\% \\
\hline
\end{tabular}
\caption{\footnotesize Comparison of the cut flows for the signal point $(\mtone,m_{\neutone})=(350,345)$ GeV in the monojet + $\met$ ATLAS search
and in the recast tool.}
\label{tab:ATLAS_mono}
\end{centering}
\end{table}

In this search a pair production of squarks is assumed in compressed scenarios where the mass difference between squark and neutralino LSP is small.
The experimental signature is characterized by an energetic jet originating from the initial state radiation and large amount of missing energy. 
The analysis performs the following pre-selection cuts:
\begin{itemize}
\item leading jet with $p_T>250$ GeV and$ |\eta|<2.4$,
\item no more than four jets with $p_T>30$ GeV,
\item lepton veto.
\end{itemize}

\begin{figure}[t]
\centering
\subfloat[]{
\label{fig:a}
\includegraphics[width=0.40\textwidth]{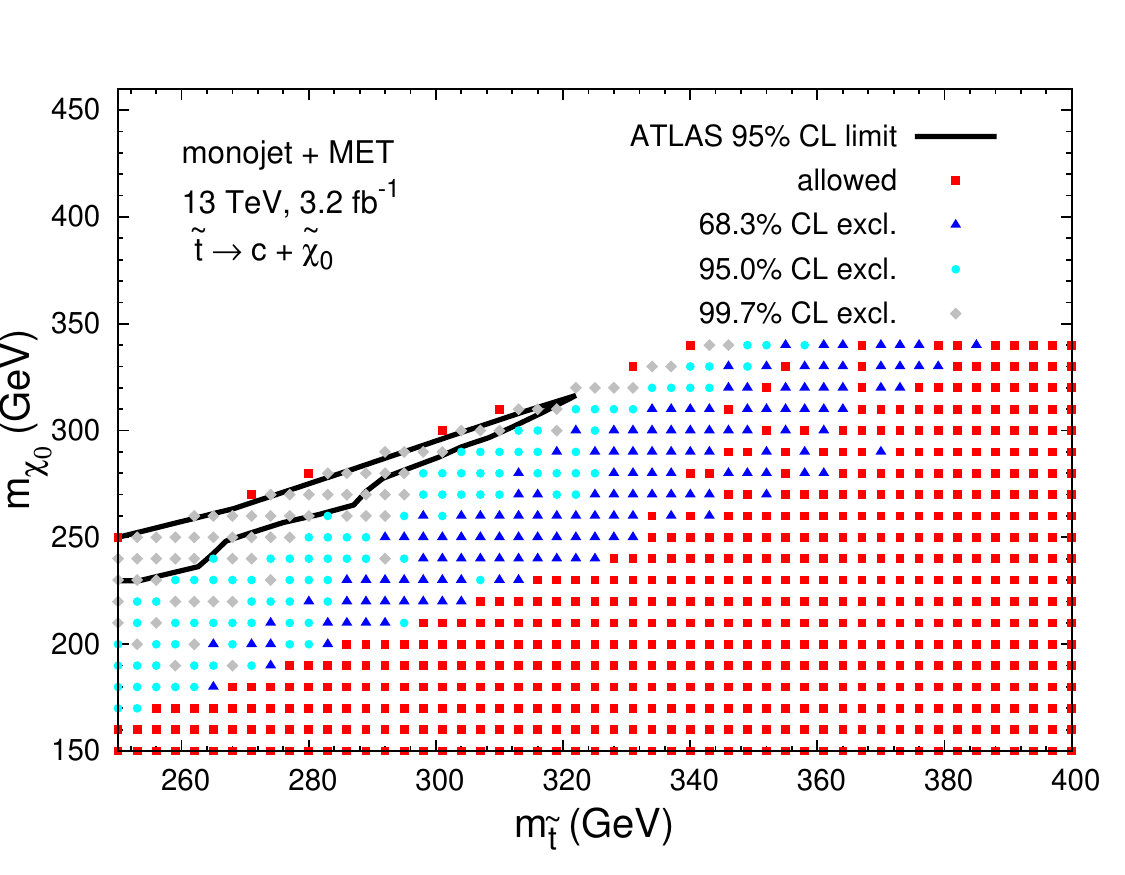}
}
\caption[]{\footnotesize \subref{fig:a} Our simulation of the ATLAS monojet search for direct stop production. 
The colour code is the same as in \reffig{fig:ATLAS_0l26j}.}
\label{fig:ATLAS_mono}
\end{figure}

Six exclusive signal bins are defined characterized uniquely by the amount of missing energy. 
In \reftable{tab:ATLAS_mono} we show a comparison of the cut flows between the experimental analysis by ATLAS and the results of our recast, 
for a signal benchmark point $(\mtone,m_{\neutone})=(350,345)$ GeV.
In \reffig{fig:ATLAS_mono} we present a validation of our simulation in terms of the exclusion limit in the parameter space of 
$(\mtone,m_{\neutone})$. Note that, despite the fact that the experimental cuts employed in this search are relatively straightforward,
our recast seems to exclude larger part of the parameter space that it was reported by ATLAS. Since the crucial element of the analysis is the 
correct identification of the initial state radiation jet, it emphasizes the importance of proper modelling of this phenomenon.

\bibliographystyle{utphysmcite}	
\bibliography{myref}


\end{document}